\documentclass[journal,twoside,web, letterpaper]{ieeecolor}
\usepackage{tac}

\usepackage{cite}[sort,compress]
\usepackage[pdftex]{graphicx}
\usepackage{array}
\usepackage{url}
\usepackage{arydshln}
\usepackage{float}
\usepackage{scrextend}
\usepackage{blindtext}
\usepackage{amsmath,amsfonts,amssymb}
\usepackage{tikz}
\usepackage{xspace}
\usepackage{xcolor}
\usepackage{mathrsfs,mathtools}
\usepackage{mathalfa}
\usepackage{euscript}
\usepackage{csquotes}
\usepackage{datetime}
\usepackage{arydshln}
\let\labelindent\relax
\usepackage[inline]{enumitem}
\usepackage[linesnumbered,ruled,vlined]{algorithm2e}

\usepackage{algpseudocode}
\algrenewcommand\algorithmicrequire{\textbf{Requires}}
\algrenewcommand\algorithmicensure{\textbf{Outputs}}

\newdateformat{monthyeardate}{\monthname[\THEMONTH], \THEYEAR}

\DeclareRobustCommand{\legendline}[1]{\hspace{-0pt}\tikz[#1,line width=0.4mm,baseline=-0.5ex]{\draw (0,0) -- (.35,0);}\hspace{-0pt}}
\DeclareRobustCommand{\legendlined}[1]{\hspace{-0pt}\tikz[#1,line width=0.4mm,baseline=-0.5ex]{\draw[dashed] (0,0) -- (.3,0);}\hspace{-0pt}}

\definecolor{mblue}{rgb}{0,0.4470,0.7410}
\definecolor{morange}{rgb}{0.8500,0.3250,0.0980}
\definecolor{myellow}{rgb}{0.9290,0.6940,0.1250}
\definecolor{mpurple}{rgb}{0.4940,0.1840,0.5560}
\definecolor{mgreen}{rgb}{0.4660,0.6740,0.1880}
\definecolor{mcyan}{rgb}{0.3010,0.7450,0.9330}
\definecolor{mred}{rgb}{0.6350,0.0780,0.1840}
\definecolor{mgreenblue}{rgb}{0.0,1.0,0.5}
\definecolor{parulablue}{rgb}{0.2431,0.1490,0.6588}
\definecolor{parulalblue}{RGB}{39,151,235}
\definecolor{parulagreen}{RGB}{129,204,89}
\definecolor{parulayellow}{RGB}{249,251,21}
\definecolor{cblue}{rgb}{0,0.9,1}
\definecolor{corange}{rgb}{1,0.7,0}
\definecolor{mgray}{rgb}{0.8,0.8,0.8}

\usepackage{amsthm}

\theoremstyle{definition}
\newtheorem{defn}{Definition}
\newtheorem{exmp}{Example}
\newtheorem{problem}{Problem}
\theoremstyle{plain}
\newtheorem{theorem}{Theorem}

\newtheorem{appendixprop}{Proposition}[subsection]
\newtheorem{lem}{Lemma}

\newtheorem{prop}{Proposition}

\newtheorem{assumption}{Assumption}
\newtheorem{condition}{Condition}
\theoremstyle{remark}
\newtheorem{rmrk}{Remark}

\newenvironment{definition}{\begin{defn}}{\hfill$\square$\end{defn}}
\newenvironment{rem}{\begin{rmrk}}{\hfill$\square$\end{rmrk}}

\newcommand{\lpvcore}{\textsc{LPVcore}\xspace}
\newcommand{\matlab}{\textsc{Matlab}\xspace}

\newcommand{\mc}[1]{\mathcal{#1}}
\newcommand{\mf}[1]{\mathfrak{#1}}
\newcommand{\mr}[1]{\mathrm{#1}}
\newcommand{\mb}[1]{\mathbb{#1}}
\newcommand{\ms}[1]{\mathscr{#1}}
\newcommand{\msf}[1]{\mathsf{#1}}
\newcommand{\mt}[1]{\mathtt{#1}}
\newcommand{\mbf}[1]{\mathbf{#1}}

\newcommand{\R}{\mathbb{R}}

\renewcommand{\P}{\mathbb{P}}
\newcommand{\dnx}{{n_\mr{x}}}
\newcommand{\dny}{{n_\mr{y}}}
\newcommand{\dnu}{{n_\mr{u}}}
\newcommand{\dnp}{{n_\mr{p}}}
\newcommand{\dna}{{n_\mr{a}}}
\newcommand{\dnb}{{n_\mr{b}}}
\newcommand{\dnw}{{n_\mr{w}}}

\newcommand{\posdef}{\succ}

\newcommand{\unaryminus}{\scalebox{0.65}[1]{\ensuremath{\,-}}}

\newcommand{\rankdef}[1]{\ensuremath{\mr{rank}\!\left(#1\right)}}

\newcommand{\rank}{\mr{rank}}

\DeclareMathOperator{\sinc}{sinc}

\newcommand{\kron}{\otimes} %
\newcommand{\bkron}{\circledcirc}

\newcommand{\Nd}{N_{\;\!\!\mr{d}}}
\newcommand{\dataset}{\mc{D}_{\Nd}}
\newcommand{\ts}{T_{\mr{s}}}

\newlength\figH 
\newlength\figW
\usepackage{pgfplots}

\newcommand{\ddict}[1]{{\breve{#1}}}
\newcommand{\inittr}[1]{#1}
\newcommand{\predtr}[1]{\bar{#1}}
\newcommand{\sigint}[3]{#1_{[#2,#3]}}
\newcommand{\psigint}[4]{#1_{[#2,#3]|#4}}
\newcommand{\Bss}{\mf{B}^{\textrm{\textsc{ss}}}}
\newcommand{\Bio}{\mf{B}^{\textrm{\textsc{io}}}}
\newcommand{\datasetss}{\dataset^{\textrm{\textsc{ss}}}}
\newcommand{\datasetio}{\dataset^{\textrm{\textsc{io}}}}

\newcommand{\Nc}{N_{\:\!\!\mr{c}}}
\newcommand{\Nkap}{N_{\tau}}
\newcommand{\termcost}{V_\mr{f}}
\newcommand{\termset}{\mb{X}_\mr{f}}
\newcommand{\LBf}{\mbf{L}(\mf{B})}
\newcommand{\eigmax}{\bar{\lambda}}
\newcommand{\eigmin}{\underline{\lambda}}

\makeatletter
\DeclareRobustCommand\vdots{%
  \mathpalette\@vdots{}%
}
\newcommand*{\@vdots}[2]{%
  \sbox0{$#1\cdotp\cdotp\cdotp\m@th$}%
  \sbox2{$#1.\m@th$}%
  \vbox{%
    \dimen@=\wd0 %
    \advance\dimen@ -3\ht2 %
    \kern.5\dimen@
    \dimen@=\wd2 %
    \advance\dimen@ -\ht2 %
    \dimen2=\wd0 %
    \advance\dimen2 -\dimen@
    \vbox to \dimen2{%
      \offinterlineskip
      \copy2 \vfill\copy2 \vfill\copy2 %
    }%
  }%
}
\DeclareRobustCommand\ddots{%
  \mathinner{%
    \mathpalette\@ddots{}%
    \mkern\thinmuskip
  }%
}
\newcommand*{\@ddots}[2]{%
  \sbox0{$#1\cdotp\cdotp\cdotp\m@th$}%
  \sbox2{$#1.\m@th$}%
  \vbox{%
    \dimen@=\wd0 %
    \advance\dimen@ -3\ht2 %
    \kern.5\dimen@
    \dimen@=\wd2 %
    \advance\dimen@ -\ht2 %
    \dimen2=\wd0 %
    \advance\dimen2 -\dimen@
    \vbox to \dimen2{%
      \offinterlineskip
      \hbox{$#1\mathpunct{.}\m@th$}%
      \vfill
      \hbox{$#1\mathpunct{\kern\wd2}\mathpunct{.}\m@th$}%
      \vfill
      \hbox{$#1\mathpunct{\kern\wd2}\mathpunct{\kern\wd2}\mathpunct{.}\m@th$}%
    }%
  }%
}
\makeatother

\newcommand{\mcG}{\mathcal{G}}   %

\newcommand{\mcV}{\mc{V}}

\newcommand{\lyap}{Z}

\def\extendedversion{1} %
\newcommand{\extver}[2]{%
  \ifx\extendedversion\undefined%
    #1%
  \else%
    #2%
  \fi%
}

\extver{%
\markboth{\journalname, VOL. XX, NO. XX, XXXX \the\year{}}
{C.~Verhoek \MakeLowercase{\textit{et al.}}: An LPV approach to DPC (\monthyeardate\today)}%
}{%
\markboth{C.~Verhoek \MakeLowercase{\textit{et al.}}: A linear parameter-varying approach to data predictive control (E\MakeLowercase{xtended version}, \monthyeardate\today)}
{C.~Verhoek \MakeLowercase{\textit{et al.}}: A linear parameter-varying approach to data predictive control (E\MakeLowercase{xtended version}, \monthyeardate\today)}%
}

\begin{document}
\title{A Linear Parameter-Varying Approach to Data Predictive Control}
\author{%
Chris Verhoek,
Julian Berberich,
Sofie Haesaert, %
Roland T{\'o}th, and
Hossam S. Abbas. %
\thanks{%
This work has received funding from the European Research Council (ERC) under the European Union's Horizon 2020 research and innovation programme (grant agreement nr. 714663),  the European Union within the framework of the National Laboratory for Autonomous Systems (RRF-2.3.1-21-2022-00002) and was also  supported by the Deutsche Forschungs-gemeinschaft (DFG, German Research Foundation) under
Project No. 419290163.
}
\thanks{C.~Verhoek, S.~Haesaert and R.~T\'oth are with the Control Systems Group, Eindhoven University of Technology, The Netherlands. J.~Berberich is with the University of Stuttgart, Institute for Systems Theory and Automatic Control, Germany. H.~S.~Abbas is with the Institute for Electrical Engineering in Medicine, Universit{\"a}t zu L{\"u}beck, Germany. R.~T\'oth is also with the Systems and Control Lab, Institute for Computer Science and Control, Hungary.
}
\thanks{Email addresses: \textbraceleft \texttt{c.verhoek}, \texttt{s.haesaert}, \texttt{r.toth}\textbraceright \texttt{@tue.nl}, \texttt{julian.berberich@ist.uni-stuttgart.de} and \texttt{h.abbas@uni-luebeck.de}. Corresponding author: C.~Verhoek.}
}

\maketitle

\begin{abstract}
	By means of the \emph{linear parameter-varying} (LPV)  Fundamental Lemma, we derive novel \emph{data-driven predictive control} (DPC) methods for LPV systems. In particular, we present output-feedback and state-feedback-based LPV-DPC methods with terminal ingredients, which guarantee exponential stability and recursive feasibility. We provide methods for the data-based computation of these terminal ingredients. Furthermore, an in-depth analysis of the application and implementation aspects of the LPV-DPC schemes is given, including application for nonlinear systems and handling noisy data. We compare and demonstrate the performance of the proposed methods {in} a {detailed} simulation example involving a nonlinear unbalanced disc system.

\end{abstract}


\section{Introduction} \label{s:intro}
\IEEEPARstart{E}{ver-increasing} performance requirements in engineering are pushing practical control design problems to become  increasingly more complex. In particular, controlling systems in operating regions with dominant nonlinear behavior is becoming important with still the need to guarantee
stability and performance {of the closed-loop operation}. A powerful framework to systematically deal with complex nonlinear control problems and obtain such guarantees is the \emph{linear parameter-varying} (LPV) framework~\cite{Toth2010_book}. In this framework, system descriptions are considered with a linear signal relation between the input, state and output signals, while this relation itself is varying along a measurable, time-varying signal $p$. The signal $p$ is referred to as the \emph{scheduling signal}, capturing nonlinear, time-varying and/or exogenous effects in the original system behavior, and hence characterizing its embedding in the solution set of {an} LPV description. Linearity of the resulting LPV surrogate representation has allowed the extension of many \emph{linear-time invariant} (LTI) control design methods, widely used in the industry, {to} address nonlinear systems.

In many engineering problems, actuator and operational constraints are required to be satisfied in closed-loop operation. For such problems, \emph{model predictive control} (MPC) is a well-suited %
{solution}
with theoretical guarantees of recursive feasibility, stability and performance~\cite{rawlings2020model}. Combining the LPV concept and MPC techniques, i.e., LPV-MPC methods such as~\cite{Toth21aIET,Casavola2008a,CiWe20}, have been shown to %
{be} advantageous %
for a large range of complex problems~\cite{morato2020lpvmpcsurvey}. However, %
often {it is} cumbersome or even impossible with first-principles modeling to obtain accurate {models in practice} %
on which these MPC methods can be deployed.  {While data-driven modeling in terms of LPV model identification, has been developed to supply reliable models for control \cite{Toth2010_book}, it still requires an elaborate toolchain and human expertise to successfully accomplish it.}
{Hence}, as an alternative, \emph{data-driven predictive control} (DPC) methods have been developed to design predictive controllers from data.

In existing \emph{LPV data-driven predictive control} (LPV-DPC) schemes, often a \emph{two-step} approach is still used under-the-hood, {which is based on an LPV model identification scheme integrated with} %
a model-based LPV-MPC design, cf.~\cite{gidon2021data, irdmousa2019data}. This two-step approach is {also applied in} \emph{continuous-time} in~\cite{luo2017data}, while the work in~\cite{cisneros2020data} uses a Koopman-based identification scheme in combination with the LPV framework to obtain the predictor. In~\cite{piga2017direct}, data-driven (predictive) controller design for SISO systems is considered using a two-component hierarchical structure, using a reference model from~\cite{formentin2016direct}. %
Data-driven predictive control in the form of learning an LPV model that is used in an LPV-MPC scheme is presented in~\cite{bao2023learning} {with formal} safety and stability guarantees. However, for all the aforementioned approaches, the quality of the identified LPV model {still} \emph{governs} the performance of the {predictive controller}, while the identification objective often significantly deviates from the control objective. {Also many choices in terms of the used model structure are still required to be accomplished by the user, similarly to regular system identification. In the LTI case, these observations} lead to the idea of designing predictive controllers \emph{directly} from data, without {the} need\footnote{{See~\cite{dorfler2023data} for {a detailed} discussion on \emph{when} and \emph{when not} to use a model.}} {for} a model identification step.

A key result in direct DPC design for LTI systems, which also allows for theoretical guarantees, is Willems' Fundamental Lemma~\cite{WillemsRapisardaMarkovskyMoor2005}. With this result, the behavior of an LTI system can be characterized using only data. The use of the Fundamental Lemma to obtain direct %
LTI-DPC schemes has been initiated by \cite{yang2015data, coulson2019deepc}, where the name `DeePC' was coined. From these works, a multitude of papers on LTI-DPC have been published, providing deeper analysis of the DPC scheme, stability/performance/robustness guarantees, and applications of the paradigm in practice~\cite{berberich2020data, verheijen2021data, vanwingerden2022data, CSMarticleDorfler}. 
DPC reaches beyond the class of LTI systems by means of (extensions of) the Fundamental Lemma with respect to linear time-periodic systems~\cite{li2022data}, Koopman-based surrogates for nonlinear systems~\cite{lian2021koopman, gholaminejad2023stable}, nonlinear systems by means of implicit online linearization~\cite{berberich2022linear} and  LPV systems~\cite{VerhoekAbbasTothHaesaert2021}. {In this paper, we build on the preliminary results of {the conference paper}~\cite{VerhoekAbbasTothHaesaert2021}, where an LPV-DPC scheme {has been} proposed for LPV systems in \emph{input-output} (IO) form. In the present work, we \emph{extend} the results of~\cite{VerhoekAbbasTothHaesaert2021} by {(i)} providing stability and feasibility guarantees for this LPV-IO-DPC scheme, {which were not present in~\cite{VerhoekAbbasTothHaesaert2021};} {(ii)} introducing {a novel} state-feedback {formulation in terms of} an LPV-SS-DPC scheme; {and} {(iii)} analyzing and extending the obtained LPV-DPC schemes to support {their} effective implementation.} %
More specifically, our contributions in this paper are:
\begin{enumerate}[label={C\arabic*:}, align=left, ref={C\arabic*}, leftmargin=*]
    \item Development of a direct data-driven output-feedback LPV-DPC scheme\footnote{{The main difference {between} the methods in this paper and LTI-DPC methods, such as DeePC~\cite{coulson2019deepc}, is the considered system class. {In fact, our work can be seen as a generalization of the DeePC method} to the class of LPV systems, meaning that the two methods coincide if the scheduling signal is taken {to be} \emph{constant} for all times (i.e., there is no parameter variation).}} {that uses only} a measured sequence of input-scheduling-output data. %
    {We show that the method} %
    is recursively feasible and guarantees constraint satisfaction and exponential stability of the closed-loop without a model or extensive prior knowledge on the system. \label{C:LPV-IO-DPC}
	\item Development of a direct data-driven state-feedback {variant of the} LPV-DPC scheme {with}  
	recursively {feasibility} 
	and %
	stability {guarantees, ensured by}
	novel terminal ingredients 
	{computed purely based on data.}
	 \label{C:LPV-SS-DPC}
	\item %
	{Providing} effective approaches for handling noise and disturbances in the data together with determining the scheduling sequence in case the {proposed DPC} approach is applied for nonlinear systems or systems dependent on exogenous effects.
    \label{C:properties}
	\item Extensive {comparison of the developed DPC schemes with nonlinear, LPV, and LTI MPC solutions, as well as a comparison with the LTI DeePC scheme.}
	 \label{C:exp}
\end{enumerate}

The paper is structured as follows: A detailed formulation of the considered problem setting is given in Section~\ref{s:problem}, while, in Section~\ref{s:dddLPVrep}, the fully data-based predictors used in the LPV-DPC schemes {are derived}. Section~\ref{s:lpv-dpc:io} provides Contribution~\ref{C:LPV-IO-DPC} by developing the LPV-IO-DPC scheme, while the LPV-SS-DPC is derived in Section~\ref{s:lpv-dpc:ss} corresponding to Contribution~\ref{C:LPV-SS-DPC}. {Analysis of the LPV-DPC schemes together with robustification against noise and computation of scheduling sequences used in the predictor} is given in Section~\ref{s:practicalproblems}, {constituting to} Contribution~\ref{C:properties}. Comparison and demonstration of the effectiveness of the approaches (Contribution~\ref{C:exp}) {are} given in Section~\ref{s:example}. {Conclusions} on the presented results are drawn in Section~\ref{s:conclusion}.

\subsubsection*{Notation}
The set of positive integers is denoted {as} $\mathbb{N}$, while $\mathbb{R}$ denotes the set of real numbers. The $p$-norm of a vector $x\in\mathbb{R}^{n_\mathrm{x}}$ is denoted by $\lVert x\rVert_p$ {and} the Moore-Penrose (right) pseudo-inverse {of a matrix} is denoted by $\dagger$. {The} Kronecker product of two matrices $A$ and $B$ {is $A\kron B$}. 
 We use $(*)$ %
 {for} a symmetric term in a quadratic expression, e.g. $(*)^{\!\top}\hspace{-1mm} A x = x^{\!\top}\hspace{-1mm} A x$ for $A\in\mathbb{R}^{n\times n}$ and $x\in\mathbb{R}^{n}$. The identity matrix of size $n$ is denoted as $I_n$ and $0_{n\times m}$ denotes the $n\times m$ zero-matrix, while $1_{n}$ denotes the vector $\begin{bsmallmatrix} 1 & \cdots & 1 \end{bsmallmatrix}^\top\in\mb{R}^n$. %
For $\mb{A}$ and $\mb{B}$, $\mb{B}^\mb{A}$ indicates the collection of all maps from $\mb{A}$ to $\mb{B}$. The projection of $\mb{D}\subseteq\mb{A}\times\mb{B}$ onto the elements of $\mb{A}$ is denoted by $\pi_a\mb{D} = \{a \in\mb{A}\mid(a,b)\in\mb{D}\}$. We denote the interior of a set by $\mr{int}(\mb{A})$.
The notation  $A \succ 0$ and $A\prec 0$ ($A \succeq0 $ and $A \preceq 0$) stands for positive/negative (semi) definiteness of $A\in\mathbb{R}^{n\times n}$. {A block-matrix of the form $\begin{bsmallmatrix}A & 0 \\ 0 & B\end{bsmallmatrix}$ is denoted by $\mr{blkdiag}(A,B)$.}
For a parameter-varying matrix $X:\mb{P}\to\mb{R}^{\dnx\times\dnx}$, we denote the maximum and minimum eigenvalue of $X(p)$ over all possible $p\in\mb{P}$ by {$\eigmax_{\mathbb{P}}(X)$} and {$\eigmin_{\mathbb{P}}(X)$}, respectively.
For a given signal $w\in(\mb{R}^{\dnw})^\mb{Z}$ and a compact set $[t_1,t_2]\subset\mb{Z}$, the notation $w_{[t_1,t_2]}$ corresponds to the truncation of $w$ to the time interval $[t_1,t_2]$.
For $w_{[1,N]}$, we denote the Hankel matrix of depth $L$ associated with it as
\begin{align*}
\mc{H}_L(w_{[1,N]}) = \begin{bsmallmatrix} 
    w_1 & w_2 & \cdots & w_{N-L+1} \\
    w_2 & w_3 & \cdots & w_{N-L+2} \\
    \vdots & \vdots & \ddots & \vdots \\
    w_{L} & w_{L+1} & \cdots & w_{N}
\end{bsmallmatrix},
\end{align*}
while the vectorization of $w_{[1,N]}$ is given as $\mr{vec}(w_{[1,N]})$. 
The block-diagonal Kronecker operator is denoted {by} $\bkron$, i.e., we have {$w_{[1,N]}\bkron I_n:=\mathrm{blkdiag}(w_1\kron I_n, \dots, w_N\kron I_n)$}. Moreover, $w_{[1,N]}^{\mt{p}}$ denotes the sequence $\{p_k\kron w_k\}_{k=1}^N$. {{Finally,} $\mathbb{I}_{\tau_1}^{\tau_2}={\left\{ \tau \in \mathbb{Z} \mid \tau_1\leq \tau \leq \tau_2 \right\}}$ {is} an index set for $\tau_1\leq \tau_2$.}

\section{Problem statement}\label{s:problem}
\subsection{System definition}\label{ss:systemdef}
Consider \emph{discrete-time} (DT) LPV systems that can be represented by \emph{state-space} (SS) representations of the form:
\begin{subequations}\label{eq:LPVSS}
\begin{align}
	x_{k+1} & = A(p_k)x_k+B(p_k)u_k, \label{eq:LPVSS:state}\\ 
	y_k & = C(p_k) x_k + D(p_k)u_k,\label{eq:LPVSS:output}
\end{align}
where $x_k\in\R^{n_\mr{x}}$, $u_k\in\R^{n_\mr{u}}$, $y_k\in\R^{{\dny}}$ and $p_k\in\mathbb{P}\subseteq\mathbb{R}^{\dnp}$ are the state, input, output and scheduling signals at time moment $k\in\mb{Z}$, respectively. The matrix functions $A:\mathbb{P}\rightarrow \mathbb{R}^{n_\mathrm{x}\times n_\mathrm{x}}$, $B:\mathbb{P}\rightarrow \mathbb{R}^{n_\mathrm{x}\times n_\mathrm{u}}$, $C:\mathbb{P}\rightarrow \mathbb{R}^{n_\mathrm{y}\times n_\mathrm{x}}$, $D:\mathbb{P}\rightarrow \mathbb{R}^{n_\mathrm{y}\times n_\mathrm{u}}$ are considered to have affine dependence\footnote{{We will discuss the generality of affine dependence in terms of LPV embeddings of nonlinear systems in Section~\ref{s:practicalproblems}.}} on $p_k$:
\begin{equation}\label{eq:LPVSSdependency}
	\begin{aligned}
	 A(p_k)&=A_0\!+\!{\textstyle\sum_{i=1}^{n_\mr{p}}}p_{k,i}A_i, \ \
	 B(p_k)=B_0\!+\!{\textstyle\sum_{i=1}^{n_\mr{p}}}p_{k,i}B_i, \\
	 C(p_k)&=C_0\!+\!{\textstyle\sum_{i=1}^{n_\mr{p}}}p_{k,i}C_i, \ \
	 D(p_k)=D_0\!+\!{\textstyle\sum_{i=1}^{n_\mr{p}}}p_{k,i}D_i,
	\end{aligned}
\end{equation}
\end{subequations}
where $\{A_i, B_i, C_i, D_i\}_{i={0}}^{n_\mr{p}}$ are real matrices with appropriate dimensions. The scheduling signal $p$ is varying in a compact, convex set $\mb{P}\subset\R^{\dnp}$. 
The solution set of \eqref{eq:LPVSS} is defined as \vspace{0.1mm}
\begin{multline}
    \mf{B} = \big\{ (u,p,x,y)\in(\R^{\dnu}\times \mathbb{P} \times \R^{\dnx}\times  \R^{\dny})^{\mathbb{Z}} \mid p\in\ms{P}\\
    \text{and \eqref{eq:LPVSS} holds } \forall k\in\mathbb{Z}  \big\}.
\end{multline}
which is called the behavior of \eqref{eq:LPVSS}. Here, $\ms{P}\subseteq \mb{P}^\mb{Z}$ corresponds to the set of admissible trajectories of $p$, e.g., rate bounded scheduling trajectories. 
We can also define the projected behaviors $\Bss = \pi_{(u,p,x)}\mf{B}$ and $\Bio = \pi_{(u,p,y)}\mf{B}$, where the latter is called the manifest or IO behavior. {It is assumed that the system has got no autonomous dynamics, which means that~\eqref{eq:LPVSS} can be chosen such that it is \emph{structurally} state observable and controllable, see~\cite{Toth2010_book}, and the associated $\Bio = \pi_{(u,p,y)}\mf{B}$ represents all possible solution trajectories of the system. Such a representation is called \emph{minimal} and, without of loss of generality, \eqref{eq:LPVSS} is chosen to be such in the rest of this paper.}
Furthermore, we consider that the LPV system \eqref{eq:LPVSS} is subject to (pointwise-in-time) input and output constraints. Hence, for all $k\in\mb{Z}$,
\begin{equation}\label{eq:input-output-constraints}
    u_k\in\mathbb{U}, \quad y_k\in\mathbb{Y},
\end{equation}
where $\mathbb{U}\subset\R^\dnu$ and $\mathbb{Y}\subset\R^\dny$ are {compact and convex} %
 input and output constraint sets.

In this work, we consider two predictive control design {scenarios} for~\eqref{eq:LPVSS}: 

\textbf{Output-feedback case:} 
For the output-feedback data-driven predictive controller design, we assume that only input-scheduling-output measurements from \eqref{eq:LPVSS} are available. These measurements are collected in the \emph{data-dictionary}
\begin{equation}\label{eq:datadictionary:io}
    \datasetio=\{\ddict{u}_k,\ddict{p}_k,\ddict{y}_k\}_{k=1}^{\Nd},
\end{equation}
where the {notation} $\ddict{\bullet}$ %
is used throughout the paper for measured signals in the data-dictionary.
To make the data-driven design tractable, we %
 {assume that the following condition is satisfied}: %
\begin{condition}\label{ass:shifted-affine-realization}
    The IO map of~\eqref{eq:LPVSS} admits a shifted-affine LPV-IO realization. That is, the manifest behavior $\Bio$ is characterized by
    \begin{subequations}\label{eq:LPVIO}
    \begin{equation}\label{eq:LPVIO:rep}
        y_k+{\textstyle\sum_{i=1}^{\dna}}\,a_i(p_{k-i})y_{k-i}={\textstyle\sum_{i=1}^{\dnb}}\,b_i(p_{k-i})u_{k-i},
    \end{equation}
    with $\dna,\dnb\geq{1}$ {and} %
    $a_i:\mb{P}\to\mb{R}^{\dny\times\dny}$ and $b_i:\mb{P}\to\mb{R}^{\dny\times\dnu}$ {that} are affine functions of the time-shifted values of %
    $p_k$, i.e., 
    \begin{align}\label{eq:LPVIO:dep}
        a_i(p_{k-i}) & = a_{i,0} + {\textstyle\sum_{j=1}^{\dnp}}a_{i,j}p_{k-i,j}, \\ 
        b_i(p_{k-i}) & = b_{i,0} + {\textstyle\sum_{j=1}^{\dnp}}b_{i,j}p_{k-i,j}.
    \end{align}
    \end{subequations}
\end{condition}
{Fulfillment of this condition is often assumed}
in %
LPV modeling {and} identification {as it enables a direct state-space realization of the IO map~\cite{toth2011state}.}
{{In} general, {an} LPV-SS representation~\eqref{eq:LPVSS} does not necessarily have an LPV-IO realization that satisfies Condition~\ref{ass:shifted-affine-realization}. 
 On the other hand, for a given LPV-IO {representation in the form of \eqref{eq:LPVIO},}  %
there always exists an affine LPV-SS representation~\eqref{eq:LPVSS}, such that their manifest behaviors coincide.}
\begin{rem}\label{rem:companion}
 \eqref{eq:LPVSS} 
 admits an IO form \eqref{eq:LPVIO}, if it has a companion observability canonical form with static-affine dependence~\cite{toth2011state}. %
    {W}hen $\dnu=\dny=1$, \eqref{eq:LPVIO} can be equivalently represented by a state-minimal~\eqref{eq:LPVSS}, where $C=\begin{bmatrix} 1 & 0 & \cdots & 0\end{bmatrix}$, $D=0$ and \vspace{-0.1mm}
    \begin{equation*}
        A(p_k) = \begin{bsmallmatrix} a_1(p_k) & 1 & & 0 \\ \vdots & & \ddots & \\ a_{{n}-1}(p_k) & 0 & & 1 \\ a_{{n}}(p_k) & 0 & \cdots & 0 \end{bsmallmatrix}, \quad B(p_k) = \begin{bsmallmatrix} b_1(p_k) \\ \vdots \\ b_{{n}}(p_k) \end{bsmallmatrix}{,}
    \end{equation*}
    where $n=\max(\dna,\dnb)$ and $a_{i}=0$ and $b_j=0$ for all $i>\dna$ and $j>\dnb$.
    When $\dnu,\dny>1$, realization of~\eqref{eq:LPVIO} in terms of~\eqref{eq:LPVSS} follows a similar scheme with the matrix coefficients $a_i$ and $b_i$ used to form the above given $A$ and $B$. Based on the independent columns of  the observability matrix of the resulting state-space form, a $T\in\mathbb{R}^{n\dny \times m}$ state-transformation can be constructed that brings the LPV-SS representation to a state-minimal form with $m\leq n\dny$ states.
\end{rem}
\textbf{State-feedback case:} 
If state measurements are directly available, then \eqref{eq:LPVSS} simplifies with $C(p)=I$ and $D(p)=0$. In this case, we consider the data-dictionary
\begin{equation}\label{eq:datadictionary:ss}
    \datasetss=\{\ddict{u}_k,\ddict{p}_k,\ddict{x}_k\}_{k=1}^{\Nd}.
\end{equation}
{In this case, output constraints such as~\eqref{eq:input-output-constraints} directly translate to state constraints, i.e., for all $k\in\mb{Z}$, $x_k\in\mb{X}:=\mb{Y}$, where $\mb{X}\subset\R^\dnx$.}
Moreover, we want to highlight that Condition~\ref{ass:shifted-affine-realization} is trivially satisfied for the state-feedback case, as
\begin{equation}\label{eq:lpvssinio}
    y_k -A(p_{k-1})y_{k-1} = B(p_{k-1})u_{k-1}, \quad y_k=x_k,
\end{equation}
is simply the time-shifted version of \eqref{eq:LPVSS:state}.

\vspace{-2mm}
\subsection{Problem formulation}
In this work, we aim to solve the purely data-driven predictive control problem for unknown LPV systems {under the}
the output-feedback and state-feedback cases. %
{More precisely}, without knowing the model of system \eqref{eq:LPVSS}, {we aim to} design a predictive controller, based on a measured data-dictionary $\dataset$, which can stabilize a desired (forced) equilibrium of~\eqref{eq:LPVSS}. %
\begin{problem}\label{problem:general}
    Consider the \emph{unknown} LPV system \eqref{eq:LPVSS} with behavior $\mf{B}$ from which an $\Nd$-length data-dictionary $\dataset$ is measured. Based {only on $\dataset$}, design a predictive controller $K_\mr{PC}$ that stabilizes the data-generating system with a priori specified performance. Furthermore, $K_\mr{PC}$ must ensure constraint satisfaction for all $k\in\mb{Z}$.
\end{problem}
In this paper, we provide a solution to this problem for both the output-feedback and state-feedback case.

\section{Data-driven LPV representations and predictors}\label{s:dddLPVrep}
For the derivation of the data-driven predictive controllers, we need the concept of data-based representations of the LPV system behavior. In this section, we will introduce such representations based on~\cite{VerhoekTothHaesaertKoch2021, Verhoek2022_DDLPVstatefb, newpaper}. These representations are instrumental for the solutions to Problem~\ref{problem:general}, i.e., Contributions~\ref{C:LPV-IO-DPC} and~\ref{C:LPV-SS-DPC}, which are presented in Sections~\ref{s:lpv-dpc:io} and~\ref{s:lpv-dpc:ss}. 
\vspace{-8mm}
\subsection{LPV Fundamental Lemma}
The data-driven representations will serve as the \emph{predictors} in the LPV-DPC schemes. {For this purpose, an} essential result is the LPV Fundamental Lemma~\cite{VerhoekTothHaesaertKoch2021}, which {provides a data-driven representation of a general LPV system. In case {Condition \ref{ass:shifted-affine-realization} is satisfied,}
the representation is directly \emph{computable}  from data and} enables the prediction of the IO response for $\Nc$ steps in the future using only the data-dictionary and a given $\Nc$-long scheduling sequence. {To compute this prediction, we take the following assumption:} %
\begin{assumption}\label{A:futureschedulingknown}
    At time step $k$, the scheduling trajectory $\sigint{p}{k}{k+\Nc{-1}}$ is known up to time step $k+\Nc{-1}$.
\end{assumption}

{Similar to other methods used for LPV prediction problems and LPV-MPC for nonlinear systems, such as~\cite{CiWe20, cisneros2018constrained, morato2019novel}, we first take this assumption for technical reasons to formulate our results. As $p$ acts as a free variable in an LPV system, we require Assumption~\ref{A:futureschedulingknown} to predict the future trajectories of the system up to a prediciton horizon. Section~\ref{ss:schedulingscenarios} shows how Assumption~\ref{A:futureschedulingknown} can be reliably overcome in practice. Similar to \emph{sequential quadratic programming} (SQP) methods, the future scheduling trajectory $\sigint{p}{k}{k+\Nc{-1}}$ is iteratively synthesized (see Algorithm~\ref{alg:iterative-p-estim} later) using an a priori selected scheduling map that defines the LPV embedding of the nonlinear system in the form of~\eqref{eq:LPVSS}.}

\begin{figure}
    \centering
    \includegraphics[width=0.9\linewidth]{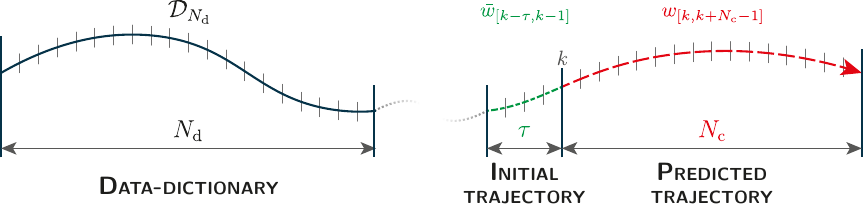}
    \caption{Prediction problem for a given data-dictionary. The signals of the system are collected here in variable $w$. {Figure adopted from~\cite{VerhoekAbbasTothHaesaert2021}}.}\label{fig:predictionproblem} \vspace{-6mm}
\end{figure}

Before we present the data-driven representations, we first introduce some required {concepts}. The \emph{state-cardinality} $\mbf{n}(\mf{B})$ {is} the minimal state-dimension among all possible LPV-SS realizations {\eqref{eq:LPVSS} of $\mf{B}$}. {We} assume {that the} LPV-SS realization {we consider to define $\mf{B}$} is minimal,  {hence $\dnx=\mbf{n}(\mf{B})$}. The \emph{lag} $\mbf{L}(\mf{B})$ is the minimum lag over all possible kernel realizations of {$\Bio$}, {i.e., $\mbf{L}(\mf{B})=\max\{\dna,\dnb\}$ if~\eqref{eq:LPVIO} is minimal. Note that $\LBf\leq\mbf{n}(\mf{B})$.} Finally, $\mf{B}|_{[t_1,t_2]}$ contains the truncation of all the trajectories in $\mf{B}$ to the interval {$[t_1,t_2]\subset \mathbb{Z}$}. {Now we can give the following key result from~\cite{newpaper}, which, in essence, gives a condition on when the data can serve as an $\Nc$-step predictor for the LPV system.

\begin{prop}[{Simplified}\footnote{{We refer to this result as the \emph{simplified} LPV Fundamental Lemma, because, under Condition \ref{ass:shifted-affine-realization}, it gives an efficiently \emph{computable} representation compared to the general LPV Fundamental Lemma of~\cite{VerhoekTothHaesaertKoch2021}. The latter does not lead to a numerically computable form as the corresponding $g$ can be potentially any meromorphic function of $\sigint{{p}}{1}{\Nc}$.}} LPV Fundamental Lemma~\cite{newpaper}]\label{prop:FLeasy}
    Given a data set $\dataset\in\Bio|_{[1,\Nd]}$ from an LPV system represented by~\eqref{eq:LPVSS} that satisfies Condition~\ref{ass:shifted-affine-realization}, and let $\Nc\geq\mbf{L}(\mf{B})$. Then, for any $\sigint{(u, p, y)}{1}{\Nc}\in\Bio|_{[1,\Nc]}${,} there exists a vector $g\in\mb{R}^{\Nd-\Nc+1}$ such that
    \begin{equation}\label{eq:prop:FLeasy}
        \hspace{-2mm}\begin{bmatrix}
        \mc{H}_{\Nc}(\sigint{\ddict{u}}{1}{\Nd}) \\
        \mc{H}_{\Nc}(\sigint{\ddict{y}}{1}{\Nd}) \\
        \mc{H}_{\Nc}(\sigint{\ddict{u}}{1}{\Nd}^{\breve{\mt{p}}}) - {\mc{P}}^{\dnu}\mc{H}_{\Nc}(\sigint{\ddict{u}}{1}{\Nd}) \\
        \mc{H}_{\Nc}(\sigint{\ddict{y}}{1}{\Nd}^{\breve{\mt{p}}}) - {\mc{P}}^{\dny}\mc{H}_{\Nc}(\sigint{\ddict{y}}{1}{\Nd})
        \end{bmatrix}\! g \!=\! \begin{bmatrix}
            \mr{vec}(\sigint{{u}}{1}{\Nc}) \\
            \mr{vec}(\sigint{{y}}{1}{\Nc}) \\
            0\\
            0
        \end{bmatrix}\!,\hspace{-1mm}
    \end{equation}
    where ${\mc{P}}^{\bullet}= \sigint{{p}}{1}{\Nc} \bkron I_\bullet$, if and only if
    \begin{equation}\label{eq:thm:dim}
        \mr{rank}\left(\begin{bmatrix}
        \mc{H}_{\Nc}(\sigint{\ddict{u}}{1}{\Nd}) \\ \mc{H}_{\Nc}(\sigint{\ddict{u}}{1}{\Nd}^{\breve{\mt{p}}}) \\ 
        \mc{H}_{\Nc}(\sigint{\ddict{y}}{1}{\Nd}) \\ \mc{H}_{\Nc}(\sigint{\ddict{y}}{1}{\Nd}^{\breve{\mt{p}}}) \end{bmatrix}\right) = \big(\dnp(\dny+\dnu) + \dnu\big)\Nc + \dnx.
    \end{equation}
\end{prop}
}
{Equation}
 \eqref{eq:thm:dim} can be considered {as a form of} %
a \emph{persistence of excitation} (PE) condition %
for LPV systems. In line with~\cite{VerhoekTothHaesaertKoch2021}, {we can see that PE}
of a data set coming from an LPV system is, next to the input signal, also dependents on the scheduling signal. As %
\eqref{eq:thm:dim} also {involves} %
the output signal {of} the data set, we could refer to it as a \emph{generalized} LPV persistence of excitation condition, similar to %
\cite{markovsky2021behavioral}. %
{In the sequel,}
we will refer to $\dataset$ satisfying%
~\eqref{eq:thm:dim} as $\dataset$ being PE of order $(\Nc,\dnx)$. {Note that~\eqref{eq:thm:dim} {also} defines the minimum number of data points, i.e., a lower bound for $\Nd$, that is required to represent the $\Nc$-length behavior of an LPV system~\eqref{eq:LPVSS} that satisfies Condition~\ref{ass:shifted-affine-realization}:}
\[ \Nd\geq\big(1+\dnp(\dny+\dnu)+\dnu\big)\Nc+\dnx-1. \]
Proposition~\ref{prop:FLeasy} can be applied for both the output-feedback case and the state-feedback case (via~\eqref{eq:lpvssinio}). Moreover, this result provides, based on the single data-sequence in $\dataset$, all possible trajectories that are compatible with any given scheduling sequence $\sigint{p}{1}{\Nc}$.  Hence, we now can formulate the data-based predictors for the output-feedback and state-feedback LPV-DPC schemes. We derive them through the 
simulation problem, cf.~\cite{MarkovskyRapisarda2008, VerhoekAbbasTothHaesaert2021, VerhoekTothHaesaertKoch2021}, depicted in Fig.~\ref{fig:predictionproblem}.
Based on an {a priori}\footnote{It is assumed that the data-dictionary $\dataset$ {was} measured offline in the past {and~\eqref{eq:thm:dim} {was} verified a posteriori. {Note that} input design for guaranteeing the satisfaction of~\eqref{eq:thm:dim} is still an open question}. {Alternatively,} the theory and schemes discussed in this paper can be modified such that the data-dictionary is updated online, which results in an adaptive scheme, {similar to the} online adaptation schemes for LTI-DPC, {such as}~\cite{berberich2022linear, lian2023adaptive}.} %
measured sequence $\dataset$, called the data-dictionary, we want to predict the continuation of {the evolution of the system trajectories} at time step $k$. For this we need %
the initial condition of the trajectory that we want to predict, which {corresponds to the previous observations of the system in the time} interval $[k-\tau,k-1]$ (a $\tau$-long window) {up to the current time step $k$.}
{We} denote these recorded signals %
{as $\sigint{u}{k-\tau}{k-1}$ (similar for $x,y,p$)}, while the predicted trajectory at $k$ {is} 
$\psigint{\predtr{u}}{0}{\Nc-1}{k}$. {We will use the {notation} $\predtr{\bullet}$ for all predicted trajectories.} Note that $u_k=\predtr{u}_{0|k}$ with this notation.

{According to Proposition~\ref{prop:FLeasy} and~\cite{newpaper}, the} initial trajectory must have a length $\tau\geq\mbf{L}(\mf{B})$. We can then {apply} \eqref{eq:prop:FLeasy} for both the initial trajectory and the predicted trajectory, such that $g$ at time step $k$ is restricted to the subspace that relates to all the predicted trajectories that are {possible continuations} of the initial trajectory.
Note that the Hankel matrices must have a depth of at least $\mbf{L}(\mf{B})+\Nc$. 
We are now ready to present the {data-driven} predictors for $\Bio$ and $\Bss$. 
\vspace{-3mm}
\subsection{Predictor formulation for $\Bio$}\label{ss:ddlpvpredictorio}
For the formulation of the predictor for the LPV-IO-DPC scheme{,} %
we %
assume that (an upper bound on) the lag of the system is known. %
Then, {along a given trajectory of $\sigint{p}{k}{k+\Nc-1}$ {that defines $\psigint{\predtr{p}}{0}{\Nc-1}{k}$,}} the following $\Nc$-step ahead predictor for the LPV-IO-DPC scheme is obtained:
\begin{equation}\label{eq:predictor:io}
\newcommand{\vspc}{\vphantom{\big)_p^d}}
    \begin{bmatrix}
        \vspc\mc{H}_{\tau}(\sigint{\ddict{u}}{1}{\Nkap}) \\
        \vspc\mc{H}_{\tau}(\sigint{\ddict{y}}{1}{\Nkap}) \\
        \vspc\mc{H}_{\tau}(\sigint{\ddict{u}}{1}{\Nkap}^{\ddict{\mt{p}}})\unaryminus\inittr{\mc{P}}_k^{\dnu}\mc{H}_{\tau}(\sigint{\ddict{u}}{1}{\Nkap}) \\ 
        \vspc\mc{H}_{\tau}(\sigint{\ddict{y}}{1}{\Nkap}^{\ddict{\mt{p}}})\unaryminus\inittr{\mc{P}}_k^{\dny}\mc{H}_{\tau}(\sigint{\ddict{y}}{1}{\Nkap}) \\
        \vspc\mc{H}_{\Nc}(\sigint{\ddict{u}}{\tau+1}{\Nd}) \\
        \vspc\mc{H}_{\Nc}(\sigint{\ddict{y}}{\tau+1}{\Nd}) \\
        \vspc\!\mc{H}_{\Nc}(\sigint{\ddict{u}}{\tau+1}{\Nd}^{\ddict{\mt{p}}})\unaryminus\predtr{\mc{P}}_k^{\dnu}\mc{H}_{\Nc}(\sigint{\ddict{u}}{\tau+1}{\Nd})\!\! \\ 
        \vspc\!\mc{H}_{\Nc}(\sigint{\ddict{y}}{\tau+1}{\Nd}^{\ddict{\mt{p}}})\unaryminus\predtr{\mc{P}}_k^{\dny}\mc{H}_{\Nc}(\sigint{\ddict{y}}{\tau+1}{\Nd})\!\! 
    \end{bmatrix}\! \!g_k \!=\!\! \begin{bmatrix}
        \vspc\!\mr{vec}(\sigint{\inittr{u}}{k\unaryminus\tau}{k\unaryminus1}) \!\!\\ 
        \vspc\!\mr{vec}(\sigint{\inittr{y}}{k\unaryminus\tau}{k\unaryminus1}) \!\!\\ 
        \vspc 0 \\
        \vspc 0 \\ 
        \vspc\!\mr{vec}(\psigint{\predtr{u}}{0}{\Nc\unaryminus1}{k}) \!\!\\ 
        \vspc\!\mr{vec}(\psigint{\predtr{y}}{0}{\Nc\unaryminus1}{k}) \!\!\\ 
        \vspc 0 \\ 
        \vspc 0
    \end{bmatrix}\!,
\end{equation}
where $\Nkap = \Nd-\Nc$, $\inittr{\mc{P}}_k^{\bullet}= \sigint{\inittr{p}}{k-\tau}{k-1} \bkron I_\bullet$ and $\predtr{\mc{P}}_k^{\bullet}= \psigint{\predtr{p}}{0}{\Nc-1}{k} \bkron I_\bullet$.  %
Note that the predicted $y$ and the required $u$ 
trajectories 
are completely determined by $g_k$, which is hence essentially the only required decision variable. Finally, note that %
for a full data-driven representation of $\Bio|_{[1,\Nc + \tau]}$, 
$\datasetio$ must be PE of order $(\Nc + \tau, \dnx)$.
\vspace{-4mm}
\subsection{Predictor formulation for $\Bss$} \label{ss:ddlpvpredictorss}
The lag of $\Bss$ is equal to 1, which can clearly be observed from~\eqref{eq:lpvssinio}. {This implies that all the information to advance the trajectory is in the {measured} $x_k=x_{0|k}$. {In terms of the} predictor, we {aim to} predict $\psigint{x}{1}{\Nc}{k}$, where  $x_{\Nc|k}$ {will be} used in the terminal ingredients. Moreover, because there is no feed-through in~\eqref{eq:lpvssinio}, $\predtr{u}_{\Nc|k}$ and $\predtr{p}_{\Nc|k}$ are obsolete, while $u_k=u_{0|k}$ itself can be taken as a decision variable. With these considerations, }
we end up with the following predictor for the LPV-SS-DPC scheme: 
\begin{equation}\label{eq:predictor:ss}
\newcommand{\vspc}{\vphantom{\big)_p^d}}
    \!\!\begin{bmatrix}
        \vspc\mc{H}_{1}(\sigint{\ddict{x}}{1}{\Nd\unaryminus\Nc}) \\
        \vspc\mc{H}_{\Nc}(\sigint{\ddict{x}}{2}{\Nd}) \\
        \vspc\mc{H}_{{\Nc}}(\sigint{\ddict{u}}{1}{{\Nd\unaryminus1}}) \\
        \vspc\mc{H}_{{\Nc}}(\sigint{\ddict{x}}{1}{{\Nd\unaryminus1}}^{\ddict{\mt{p}}})\unaryminus\predtr{\mc{P}}_k^{\dnx}\mc{H}_{{\Nc}}(\sigint{\ddict{x}}{1}{{\Nd\unaryminus1}}) \!\\
        \vspc\mc{H}_{{\Nc}}(\sigint{\ddict{u}}{1}{{\Nd\unaryminus1}}^{\ddict{\mt{p}}})\unaryminus\predtr{\mc{P}}_k^{\dnu}\mc{H}_{{\Nc}}(\sigint{\ddict{u}}{1}{{\Nd\unaryminus1}}) \!
    \end{bmatrix}\! g_k \!=\! \begin{bmatrix}
        \vspc x_k \\ 
        \vspc\!\mr{vec}(\psigint{\predtr{x}}{1}{\Nc}{k})\!\! \\ 
        \vspc\!\mr{vec}(\psigint{\predtr{u}}{0}{{\Nc\!\unaryminus1}}{k})\!\! \\ 
        \vspc 0 \\ 
        \vspc 0
    \end{bmatrix}\!,
\end{equation}
where $\predtr{\mc{P}}_k^{\bullet}= \psigint{\predtr{p}}{0}{{\Nc-1}}{k} \bkron I_\bullet$ and $\datasetss$ must be PE of order $(\Nc+1, \dnx)$. {As with~\eqref{eq:predictor:io}, the predicted $\predtr{x}$ and the required $\predtr{u}$ trajectories 
are completely determined by $g_k$.} %

\vspace{-1mm}
\section{LPV-DPC with input-output measurements} \label{s:lpv-dpc:io}
With the data-driven predictors defined, we now present the solution to Problem~\ref{problem:general} for the {output-feedback} case. We formulate the LPV-IO-DPC scheme using only a PE data set $\datasetio$ from an unknown LPV system {and} show that we can get guarantees on stability and recursive feasibility under mild conditions. {We want to highlight that this section {generalizes} the preliminary results {of}~\cite{VerhoekAbbasTothHaesaert2021}. In contrast to~\cite{VerhoekAbbasTothHaesaert2021}, we provide an LPV-IO-DPC scheme that guarantees recursive feasibility, constraint satisfaction and closed-loop exponential stability.}

\vspace{-2mm}
\subsection{LPV-IO-DPC scheme}
We formulate the LPV-IO-DPC scheme in a regulation scenario for a IO setpoint reference $(u^\mr{r}, p^\mr{r}, y^\mr{r})$. Because we work with IO data and thus do not have access to the state, it is difficult to formulate the LPV-IO-DPC scheme with %
terminal ingredients that are based on
an internal {$x$} associated with the trajectories in $\Bio$. Therefore, we follow the lines of~\cite{berberich2020data} by considering a terminal equality constraint that can guarantee exponential stability of the closed-loop. For this reason, {we consider the setpoint to be a forced equilibrium point of $\Bio$:}
\begin{definition} %
\label{def:equilibriumpoint}
    {A} %
    $(u^\mr{r}, p^\mr{r}, y^\mr{r})\in\mb{U}\times\mb{P}\times\mb{Y}$ is {a forced} equilibrium of %
    $\Bio$, if $\{u_k, p_k, y_k\}_{k=1}^{\LBf+1}$ with $(u_k, p_k, y_k)=(u^\mr{r}, p^\mr{r}, y^\mr{r})$ for all {$k \in \mathbb{I}_1^{\LBf+1}$}
     is a trajectory of $\Bio$. 
\end{definition}
{In Remark~\ref{remark:artificialeq}, we will discuss how to obtain this setpoint equilibrium for an unknown system directly from data.}
For {an} equilibrium $(u^\mr{r}, p^\mr{r}, y^\mr{r})$, we denote {by} $u^\mr{r}_n$ the {stacked} 
column vector containing $n$ times $u^\mr{r}$, similarly for $p^\mr{r}_n, y^\mr{r}_n$. Note that for our LPV-IO-DPC scheme, Assumption~\ref{A:futureschedulingknown} 
{implies}
that {at $k\in\mathbb{Z}$}, we {have} %
$\psigint{p}{{-\tau}}{{\Nc-1}}{k}$ available {(with $\psigint{\inittr{p}}{{-\tau}}{{-1}}{k}=\sigint{p}{k-\tau}{k-1}$ 
corresponding to the measured past of~$p$ 
and $\psigint{\predtr{p}}{{0}}{{\Nc-1}}{k}$ {being its assumed future})}. %
The cost function used for the LPV-IO-DPC scheme is %
as follows:
\begin{equation*}
   J_{\Nc}({\sigint{(u,p,y)}{k-\tau}{k-1}}, \psigint{\predtr{p}}{{0}}{{\Nc-1}}{k}, g_k) = {\textstyle\sum_{i={0}}^{{\Nc-1}}}\ell(\predtr{u}_{i|k}, \predtr{y}_{i|k}),
\end{equation*}
where %
$\ell$ %
{is} the quadratic stage cost that penalizes the distance w.r.t. the setpoint equilibrium:
\[ 
    \ell(\predtr{u}_{i|k}, \predtr{y}_{i|k}) = (*)^\top Q (\predtr{y}_{i|k}-y^\mr{r}) + (*)^\top R (\predtr{u}_{i|k}-u^\mr{r}), 
\]
with $Q\posdef0, R\posdef0$ {being performance} tuning matrices, similar to those in LQR control design. %
With this, we propose the %
LPV-IO-DPC scheme with terminal equality constraints {as}
\begin{subequations}\label{eq:lpvdpc_opt:io}
\begin{align}
    \min_{g_k}  \quad & {\textstyle\sum_{i={0}}^{{\Nc-1}}}\,\ell(\predtr{u}_{i|k},\predtr{y}_{i|k}) \label{eq:lpvdpc_opt:io:cost}\\
    \text{s.t.} \quad & \text{\eqref{eq:predictor:io} {holds}}, \label{eq:lpvdpc_opt:io:rep}\\
    & \predtr{u}_{i|k}\in\mb{U}, \ \predtr{y}_{i|k}\in\mb{Y},  \ {\forall i \in \mathbb{I}_0^{\Nc-1}},\\
    & \!\!\begin{bmatrix} u^\mr{r}_{\tau} \\
    y^\mr{r}_{\tau} \end{bmatrix} =\begin{bmatrix} \mr{vec}(\psigint{\predtr{u}}{{\Nc-\tau}}{{\Nc-1}}{k}) \\
    \mr{vec}(\psigint{\predtr{y}}{{\Nc-\tau}}{{\Nc-1}}{k})\end{bmatrix}, \label{eq:lpvdpc_opt:io:term}
\end{align}
\end{subequations}
where $\tau\geq\LBf$, and~\eqref{eq:lpvdpc_opt:io:term} are the terminal equality constraints {that force the stage cost to $0$ in the last $\tau$ steps of the control horizon. This guarantees stability and  recursive feasibility, and is common in MPC~\cite[Sec.~2.5.6]{rawlings2020model}.}
{Note that, in terms of~\eqref{eq:predictor:io}, $\predtr{u}_{i|k},\predtr{y}_{i|k}$ are completely determined by~$g_k$, which is the decision variable in \eqref{eq:lpvdpc_opt:io}.}
Problem~\eqref{eq:lpvdpc_opt:io}, is solved as a \emph{quadratic program} (QP) and applied in a receding horizon fashion (see Step \ref{alg:input:app} in Algorithm~\ref{alg:lpvdpc:io}).
\begin{algorithm}[t]\DontPrintSemicolon
\caption{%
LPV-DPC {under output-feedback}}\label{alg:lpvdpc:io}
\begin{algorithmic}[1]
\State \textbf{initialization}: set $k \gets k_0$ (starting time)
\Loop
   \State \textbf{measure} {$(u_{[k-\tau, k-1]}, p_{[k-\tau, k-1]}, y_{[k-\tau, k-1]}$)} %
   \State {\textbf{get} $\psigint{\predtr{p}}{0}{\Nc-1}{k}$\label{alg:lpvdpc:io:start}}
   \State \textbf{solve} {the QP} \eqref{eq:lpvdpc_opt:io}
  \State  \textbf{apply}  $u_k=\predtr{u}_{0|k}$ \label{alg:input:app}
  \State  \textbf{set} $k \gets k+1$ %
  \EndLoop
    \end{algorithmic} 
\end{algorithm}
{By comparing~\eqref{eq:lpvdpc_opt:io} to LTI-DPC (e.g., DeePC), the data-driven representation in~\eqref{eq:lpvdpc_opt:io:rep} is varying along the scheduling signal, representing an LPV behavior that is richer than the one in DeePC, where the behavior is inherently restricted to be LTI. Hence, \eqref{eq:lpvdpc_opt:io} can be seen as a generalization of the celebrated DeePC scheme.}

\vspace{-3mm}
\subsection{Stability and recursive feasibility}

Let us for the remainder of this section abbreviate the optimal cost at time step $k$ as $J^*_{\Nc}(k)$ 
\[ J_{\Nc}^*(k)\coloneqq J_{\Nc}({\sigint{(u,p,y)}{k-\tau}{k-1}},\psigint{\predtr{p}}{{0}}{{\Nc-1}}{k}, g_k^\ast),\]
where  $g_k^\ast$ is the minimizer of \eqref{eq:lpvdpc_opt:io}. 
For notational brevity, we assume %
 $(u^\mr{r}, y^\mr{r})=(0,0)$, which is without loss of generality due to linearity of LPV systems.
{Moreover, let $x_k$ be the state of the unknown minimal LPV-SS representation \eqref{eq:LPVSS} of the data-generating system, which is induced by the initial trajectory $\sigint{(u,p,y)}{{k-\tau}}{{k-1}}
$}. To show recursive feasibility and exponential stability of the proposed LPV-IO-DPC scheme, we require the following {standard} weak controllability condition ({see}~\cite[Assum.~2.17]{rawlings2020model}).
\begin{condition}\label{A:IO_upper_bound}
    The optimal cost {$J_{\Nc}^*(k)$} of %
    ~\eqref{eq:lpvdpc_opt:io} is quadratically upper bounded, i.e.,
    there exists a $c_{\mathrm{u}}>0$ such that, for any 
    initial condition $x_k$ corresponding to {the feasible trajectory} {$\sigint{(u,p,y)}{k-\tau}{k-1}$}, it holds that
    \begin{align}
        J_{\Nc}^*(k) 
    \leq c_{\mathrm{u}}\lVert x_k\rVert^2_{2}.
    \end{align}
\end{condition}
{Note that Condition \eqref{A:IO_upper_bound} is satisfied under convex polytopic constraints as the resulting optimal cost of {P}roblem~\eqref{eq:lpvdpc_opt:io} is continuous and piece-wise quadratic~\cite{bemporad2002explicit}.}

{One of the main technical challenges {for proving} exponential stability and recursive feasibility of the LPV-IO-DPC scheme is {characterizing and ensuring} LPV input-output-to-state-stability (IOSS), which is an extension of the LTI {result} in~\cite{cai2008input}. We have summarized the solution to this problem in the following lemma:}

\begin{lem}\label{lemma:detect}
    Consider an LPV system  {defined by a state-minimal  \eqref{eq:LPVSS}} with behavior $\mf{B}$. Then, {$\mf{B}$}  is controllable and there exists an IOSS Lyapunov function ${W}(x_k,p_k)=x_k^\top P(p_k)x_k$ with $P(\msf{p})\posdef0$ for all $\msf{p}\in\P$ {such} that{, for suitable $c_1,c_2>0$,} %
    \begin{multline}\label{eq:lmdetect:bound}
        {W}(x_{k+1},p_{k+1}) - {W}(x_k, p_k) \le -\tfrac{1}{2}\|x_k\|_2^2  \\ + c_1\|u_k\|_2^2+ c_2\|y_k\|_2^2,
    \end{multline}
   {for all $(u,p,x,y)\in\mf{B}$ and $k\in\mathbb{Z}$}.
\end{lem}
See Appendix~\ref{appendix:ioss} for the proof. %
The following result shows that the proposed LPV-IO-DPC scheme controlling~\eqref{eq:LPVSS} is recursively feasible and ensures
closed-loop constraint satisfaction as well as exponential stability.

\begin{theorem}%
    Given a $\datasetio$ {from the data-generating system defined by a state-minimal~\eqref{eq:LPVSS} with behavior $\mf{B}$. Let $\datasetio$ be} PE of order $(\Nc+\tau,\dnx)$ with $\tau\geq\LBf$. If Condition~\ref{A:IO_upper_bound} is satisfied and the LPV-IO-DPC problem \eqref{eq:lpvdpc_opt:io} is feasible at $k_0$, then for all $p\in\ms{P}$ and $k>k_0$
    \begin{enumerate}[label={(\roman*)}, ref={(\roman*)}]
        \item the LPV-DPC problem \eqref{eq:lpvdpc_opt:io} is feasible,
        \item the closed-loop system satisfies the constraints: $u_k\in\mb{U}, y_k\in\mb{Y}$, \label{thm:io:cons}
        \item $(u^\mr{r}, y^\mr{r})=(0,0)$ is {an} exponentially stable {equilibrium of} the closed-loop {system}.
    \end{enumerate}
\end{theorem}
\begin{proof}
    In order to prove recursive feasibility (i), we define a feasible candidate input at time $k+1$ by shifting the previously optimal solution and appending  {it with} $0$, i.e.,
    ${u}_{i|k+1}={u}^\ast_{i+1|k}$, {$i\in \mathbb{I}_0^{\Nc-2}$} and ${u}_{\Nc-1|k+1}=0$.
{This input and the corresponding output trajectory satisfy the constraints of Problem~\eqref{eq:lpvdpc_opt:io}. Moreover, by Proposition~\ref{prop:FLeasy} and the PE assumption, there exists a variable $g_{k+1}$ such that all constraints of Problem~\eqref{eq:lpvdpc_opt:io} are fulfilled. This also implies constraint satisfaction (ii).}

For showing exponential stability, {we use standard sub-optimality arguments, cf.~\cite{rawlings2020model}, combined with the result of Lemma~\ref{lemma:detect}.} The above-defined candidate solution implies \vspace{-0.1mm}
\begin{align}\label{eq:thm_IO_lyap_decay}
    J_{\Nc}^*(k+1)-J_{\Nc}^*(k)\leq -\ell(\predtr{u}_{0|k},\predtr{y}_{0|k}).
\end{align}
Lemma~\ref{lemma:detect} implies the existence of an \emph{input-output-to-state stability} (IOSS) Lyapunov function $W(x_k, p_k)=x_k^\top P(p_k) x_k$, $P(\msf{p})\succ0$ for all $\msf{p}\in\mb{P}$ {and $k\geq0$} satisfying \eqref{eq:lmdetect:bound} 
with $c_1,c_2>0$ defined as in \eqref{eq:lmdetect:const}. With $W$, we can define the Lyapunov function candidate $V(k)= J_{\Nc}^*(k)+\gamma W(x_k, p_k)$ for some $\gamma>0$, which will show exponential stability of the closed-loop system. Note that,  {under Condition}~\ref{A:IO_upper_bound}, $V$ has trivial quadratic lower and upper bounds for all feasible $x_k$: 
\[
    \gamma\eigmin(P(p_k))\|x_k\|_2^2\le V(k)\le \big(c_\mr{u} + \gamma\eigmax(P(p_k))\big)\|x_k\|_2^2.
\]
Combining~\eqref{eq:thm_IO_lyap_decay} and~\eqref{eq:lmdetect:bound} and choosing $\gamma=\frac{\eigmin(Q,R)}{\max\{c_1,c_2\}}$, we obtain
\(%
    V(k+1)-V(k)\leq-\gamma\lVert x_k\rVert_2^2.
\) %
Then it follows from standard Lyapunov arguments with Lyapunov function $V$ that the origin of the closed-loop system is exponentially stable. 
\end{proof}
The proof of this result {follows the} line of reasoning in the proof of the LTI case, see \cite[Thm.~2]{berberich2020data}. This is possible due to linearity {of the} LPV system along a scheduling signal~$p$. The main difference compared to the LTI case, is that the detectability argument used in \cite[Thm.~2]{berberich2020data} is needed to be recast for the LPV case, which requires the {nontrivial} technical derivations of Lemma~\ref{lemma:detect}.
As the considered realization of the data-generating LPV system \eqref{eq:LPVIO} admits a {stabilizable state-}minimal LPV-SS representation that has a static-affine scheduling dependence~\cite{toth2011state, petreczky2023minimal}, we could exploit this property to formulate an IOSS Lyapunov function for the LPV system. As in \cite{berberich2020data}, the LPV IOSS Lyapunov function is used to prove stability of the origin of the closed-loop. %
\begin{rem}\label{remark:artificialeq}
    In practice, the setpoint equilibrium $(u^\mr{r}, p^\mr{r}, y^\mr{r})$ is often not given in full, but usually only in terms of {a} desired output {at a given scheduling}, e.g., a desired speed (output) for a given altitude (scheduling) in an aircraft control {problem}. A possible method to find the corresponding $u^\mr{r}$, such that $(u^\mr{r}, p^\mr{r}, y^\mr{r})$ satisfies Definition~\ref{def:equilibriumpoint}, is to use Proposition~\ref{prop:FLeasy} %
    to compute $u^\mr{r}$ via, e.g., the quadratic program:
    \begin{align*}
        u^\mr{r} = &\arg\min \|g\|_2^2 \text{\ \ {subject to}} \\
        &\begin{bmatrix}
        \mc{H}_\tau(\sigint{\ddict{u}}{1}{{\Nd-1}}) \\
        \mc{H}_\tau(\sigint{\ddict{y}}{{2}}{\Nd}) \\
        \mc{H}_\tau(\sigint{\ddict{u}}{1}{{\Nd-1}}^{\breve{\mt{p}}}) - \mc{P}^{\dnu}\mc{H}_\tau(\sigint{\ddict{u}}{1}{{\Nd-1}}) \\
        \mc{H}_\tau(\sigint{\ddict{y}}{{2}}{\Nd}^{\breve{\mt{p}}}) - \mc{P}^{\dny}\mc{H}_\tau(\sigint{\ddict{y}}{{2}}{\Nd})
        \end{bmatrix}\! g \!=\! \begin{bmatrix}
            1_\tau \kron u^\mr{r} \\
            1_\tau \kron y^\mr{r} \\0\\0
        \end{bmatrix},
    \end{align*}
    where $\mc{P}^{\bullet} = I_\tau\kron(p^\mr{r}\kron I_\bullet)$, or using 
    {an artificial equilibrium as in~\cite{limon2008mpc,berberich2020IFACchangingsetpoints}.}
     {Alternatively, the need for $u^\mr{r}$ can be alleviated by formulating a so-called offset-free cost~\cite{verheijen2021data} in terms of $\Delta\predtr{u}_{i|k} := \predtr{u}_{i|k}-\predtr{u}_{i-1|k}$ and have $\Delta\psigint{\predtr{u}}{\Nc-\tau}{\Nc-1}{k} = 0$ as terminal equality constraint.}
\end{rem}
\begin{rem}
    We want to highlight that in the LTI case it is possible to avoid the use of terminal equality constraints, and formulate an LTI-IO-DPC scheme with a terminal cost and a terminal set constraint by defining the extended state vector
    \( \xi_k = \begin{bmatrix} y_{k-1}^\top & \cdots & y_{k-\dna}^\top & u_{k-1}^\top & \cdots & u_{k-\dnb}^\top \end{bmatrix}^\top  \),
    and 
    {computing}
    the terminal cost and terminal set using the data-driven representation of the associated state-space realization. {For the LPV case, this is also possible when the IO representation~\eqref{eq:LPVIO} is considered with coefficients that are \emph{statically} dependent on $p_k$, i.e., $a_i(p_k)$ and $b_i(p_k)$. In that case, the LPV-IO representation has a nominal LPV-SS realization with static dependence as it is shown in \cite{toth2011state}. This gives an alternative solution of the LPV-DPC problem under output-feedback {and, in principle, would provide a similar DPC scheme that we will discuss in Section~\ref{s:lpv-dpc:ss}, see also~\cite{berberich2021terminal}.} However, multiple problems exist with this formulation: (i) non-minimality of this representation makes it hard to satisfy the PE condition in practice~\cite{spin2024unified} and (ii) the corresponding data-driven driven representation grows much faster in size with increasing $\Nc$ than in the LTI case, due to the lower two blocks in \eqref{eq:predictor:ss}.}     
\end{rem}

\section{LPV-DPC with state measurements} \label{s:lpv-dpc:ss}
In this section, we propose the LPV-SS-DPC scheme %
to solve Problem~\ref{problem:general} for the \emph{state-feedback} case. Again, we show that the proposed DPC scheme guarantees recursive feasibility, exponential stability and constraint satisfaction of the closed-loop operation, while only using a data set $\datasetss$ from an unknown LPV system.
\vspace{-3mm}
\subsection{LPV-SS-DPC scheme}\label{ss:schemess}
We formulate the LPV-SS-DPC scheme for regulation of a setpoint reference $(u^\mr{r}, p^\mr{r}, x^\mr{r})$, which is not required to be an equilibrium point, contrary to the output-feedback case. {The input setpoint can be seen as, e.g., an a priori known feedforward signal. Alternatively, its value can also be directly synthesized by the predictive control problem, see  Remark~\ref{remark:artificialeq}.} Consider time moment $k\in\mathbb{Z}$ and, in line with the state-feedback {setting}, $x_k$ measured from the system,  representing its current state. To drive $x_k$ to the setpoint, we consider the cost function $J_{\Nc}$ of the  %
LPV-SS-DPC scheme %
{as:}
\begin{equation}\label{eq:OLcost:ss}
    J_{\Nc}(x_k, {p_k}, \psigint{\predtr{p}}{1}{\Nc-1}{k}, g_k) = {\textstyle\sum_{i=0}^{\Nc-1}}\ell(\predtr{u}_{i|k},\predtr{x}_{i|k}),
\end{equation}
where for $\ell$, %
a quadratic stage cost that penalizes the distance w.r.t. the setpoint reference {is chosen:} %
\begin{equation}\label{eq:quadOLcost:ss}
    \ell(\predtr{u}_{i|k},\predtr{x}_{i|k}) = (*)^\top Q (\predtr{x}_{i|k}-x^\mr{r}) + (*)^\top R (\predtr{u}_{i|k}-u^\mr{r}),
\end{equation}
with $Q, R\posdef 0$. 
Note that $J_{\Nc}$ does not depend on $\predtr{u}_{i|k},\predtr{x}_{i|k}$, as these are implicitly defined by $g_k$ through the predictor~\eqref{eq:predictor:ss}. 
In general, it is well-known that directly implementing {a predictive control} scheme in terms of minimization of~\eqref{eq:OLcost:ss} does not ensure recursive feasibility, and can even destabilize an already stable system.
Therefore, we propose an LPV-SS-DPC scheme for $\Bss$ with %
a terminal cost $\termcost$ and a terminal set $\termset$, which allow to 
prove closed-loop stability and recursive feasibility of the optimization problem:
\begin{subequations}\label{eq:lpvdpc_opt:ss}
\begin{align}
    \min_{g_k}  &\quad  \termcost(\predtr{x}_{\Nc|k}-x^\mr{r})+{\textstyle\sum_{i={0}}^{\Nc-1}}\ell(\predtr{u}_{i|k},\predtr{x}_{i|k}) \label{eq:lpvdpc_opt:ss:cost}\\
    \text{s.t.} &\quad  \text{\eqref{eq:predictor:ss} holds with } {\predtr{x}_{0|k}=x_k}, \\
    & \quad \predtr{u}_{i|k}\in\mb{U}, \ \predtr{x}_{i+1|k}\in\mb{X}, \  {{\forall}i\in\mathbb{I}_0^{\Nc-1}}, \label{eq:lpvdpc_opt:ss:input} \\ %
    & \quad \predtr{x}_{\Nc|k}\in\termset. \label{eq:lpvdpc_opt:ss:term}
\end{align}
\end{subequations}
As in standard model-based predictive control, \eqref{eq:lpvdpc_opt:ss} is solved in a receding horizon fashion, as summarized in Algorithm~\ref{alg:lpvdpc:ss}.
\begin{algorithm}[t]\DontPrintSemicolon
\caption{LPV-DPC under state-feedback}\label{alg:lpvdpc:ss}
\begin{algorithmic}[1]
\State \textbf{initialization}: set $k \gets k_0$ (starting time)
\Loop
   \State \textbf{measure} $x_k$ and $p_k$ \label{alg:lpvdpc:ss:start} 
   \State {\textbf{get} $\psigint{\predtr{p}}{1}{{\Nc-1}}{k}$}
   \State \textbf{solve} {the QP} \eqref{eq:lpvdpc_opt:ss}
  \State  \textbf{apply}  $u_k=\predtr{u}_{0|k}$
  \State  \textbf{set} $k \gets k+1$ %
  \EndLoop
    \end{algorithmic} 
\end{algorithm}
In order to prove stability %
of the proposed DPC scheme,
it must be ensured that the terminal set $\termset$ is \emph{positively invariant} (PI). 
The %
conditions on $\termset$ and $\termcost$ %
that allow to prove stability and recursive feasibility will be discussed in the next section. 

\vspace{-3mm}
\subsection{Stability and recursive feasibility}\label{ss:stab_ss}

In this section, we prove recursive feasibility of the optimization problem \eqref{eq:lpvdpc_opt:ss} as well as exponential stability of the data-generating system $\mf{B}^\textrm{\textsc{ss}}$ under the proposed LPV-SS-DPC control law. {Again, }%
we  assume  {without loss of generality} %
that the setpoint reference is $(x^\mr{r},u^\mr{r})=(0,0)$.
{Let us} denote the optimal cost of \eqref{eq:lpvdpc_opt:ss} by $J_{\Nc}^*(x_k,p_k, \psigint{\predtr{p}}{1}{\Nc-1}{k})$ {and} %
define the closed-loop state transition map {of \eqref{eq:LPVSS} with the LPV-DPC scheme \eqref{eq:lpvdpc_opt:ss}} by $\phi_{\mr{cl}}$ such that $x_{k+1}=\phi_{\mr{cl}}(x_k,p_k)$. 

We assume that the following conditions hold, which are standard when considering 
terminal ingredients, {see}, e.g.,~\cite[Assum.~2.14]{rawlings2020model}. {We comment on satisfying {these conditions} using only~$\datasetss$ in %
Section~\ref{ss:TerComp}.}
\begin{condition}\label{A:termingredients}    
    The following {conditions} are satisfied:
    \begin{enumerate}[label=\alph*), ref={\alph*}]
        \item The data-generating system with behavior $\Bss$ is quadratically stabilizable, i.e., there exist a positive definite function $\termcost(x) = x^{\!\top} \hspace{-0.5mm} \lyap x$, where $\lyap \succ 0$, and an associated control law $u=K(p)x$ such that
        \begin{equation}\label{eq:qstab}
            \termcost\left(\phi_{\mr{cl}}(x,p)\right) - \termcost(x) \leq  -\|x\|_{Q+K^\top \! (p)RK(p)}^2,
        \end{equation}
        for all $x\in\pi_x\Bss, p\in\ms{P}$. \label{A:termingredients:qstab}
        \item The set $\mathbb{X}_\mr{f}\subset\mathbb{X}$ is PI for the data-generating system~\eqref{eq:LPVSS} with behavior $\Bss$ {under} ${u}=K({p}){x}$, i.e., $\phi_\mr{cl}(\termset,\mb{P})\subseteq\termset$ and %
        $0\in\mathrm{int}(\termset)$.
    \end{enumerate}
\end{condition}
If the system with behavior $\mf{B}^\textrm{\textsc{ss}}$ is quadratically stabilizable, then {there exists a state feedback law $K(p)$ such that}
$x^\mr{r}=0$
is globally exponentially stable for the resulting closed-loop system $\forall p^\mr{r}\in\mathbb{P}$. {For $x^\mr{r}\neq0$, global exponential stability is guaranteed if $p_k\to p^\mr{r}$ {sufficiently fast}.} {We will show in Section \ref{ss:TerComp}  that, for Condition \ref{A:termingredients}, if a stabilizing $K(p)$ exists, then it can be directly designed based on only $\datasetss$.}

The following result shows the desired recursive feasibility and stability properties of the closed-loop under Algorithm~\ref{alg:lpvdpc:ss}.
\begin{theorem}[LPV-SS-DPC recursive feasibility and exponential stability] \label{thm:lpvssdpc}
    Given a $\datasetss$ {from the data-generating system with behavior $\Bss$} that is PE of order $(\Nc+1,\dnx)$. If %
    Condition~\ref{A:termingredients} is satisfied and the LPV-SS-DPC problem \eqref{eq:lpvdpc_opt:ss} is feasible at $k_0$, then for all $p\in\ms{P}$ and $k>k_0$
    \begin{enumerate}[label={(\roman*)}, ref={(\roman*)}]
        \item the LPV-SS-DPC problem \eqref{eq:lpvdpc_opt:ss} is feasible,\label{thmitem:feas}
        \item the closed-loop system satisfies the constraints, i.e., $u_k\in\mb{U}, x_k\in\mb{X}$, \label{thmitem:cons}
        \item {the origin of $\Bss$ is an {exponentially} stable equilibrium of the closed-loop system}. \label{thmitem:stab}
    \end{enumerate}
\end{theorem}
\begin{proof}
    See Appendix~\ref{appendix:pf:thmlpvssdpc}.
\end{proof}

The {proof of Theorem \ref{thm:lpvssdpc}} follows the same line of reasoning as that of standard MPC stability theory, thanks to the existence of the data-driven predictor~\eqref{eq:predictor:ss} and solvability of the data-based state-feedback design problem (see Section \ref{ss:TerComp}). {Hence, the technical challenges for this result lie in the efficient formulation of~\eqref{eq:predictor:ss}, and in the computation of the terminal ingredients in a fully data-driven setting.}

\subsection{Computation of terminal components}\label{ss:TerComp}
In this section, we give %
methods that allow {the construction of the terminal ingredients of} the LPV-SS-DPC {approach by computationally efficient linear or quadratic programs} using only the data set $\datasetss$. %
\subsubsection{Computation of {$\mathit{K(p)}$} and $\mathit{\termcost}$}
We first discuss the design of a terminal state-feedback controller {$K(p)$} that satisfies Condition~\ref{A:termingredients}.\ref{A:termingredients:qstab}.
We consider %
{$K(p)$} as an
LPV controller with affine scheduling dependence: %
\begin{equation}\label{eq:fblaw}
    u_k = K(p_k)x_k, \quad K(p_k) = K_0 + {\textstyle\sum_{i=1}^{\dnp}}p_{k,i}K_i.
\end{equation}
{In} %
\cite[Thm.~4]{Verhoek2022_DDLPVstatefb}, %
a direct data-driven method for the design of {stabilizing state-feedback} controllers %
{has been derived.}
We give here a brief summary {of it}, and refer to~\cite{Verhoek2022_DDLPVstatefb} for further %
 details. {Alternatively, data-driven synthesis of terminal ingredients using set-theoretical approaches such as in~\cite{MillerSznaier2022, verhoek2024decoupling} can be used.} 
 
 For the synthesis, we only need $Q,R$ {(defining the quadratic performance \eqref{eq:OLcost:ss})} and a data dictionary $\datasetss$ that satisfies {the simplified form of the PE condition~\eqref{eq:thm:dim} in the state-feedback case, i.e., when $y_k=x_k$ and $\Nc=1$:}
\begin{equation}\label{eq:rkcond}
    \rank\left(\begin{bmatrix} \mc{H}_1(\sigint{\ddict{x}}{1}{\Nd-1} \\ \mc{H}_1(\sigint{\ddict{x}}{1}{\Nd-1}^{\ddict{\mt{p}}} \\ \mc{H}_1(\sigint{\ddict{u}}{1}{\Nd-1} \\ \mc{H}_1(\sigint{\ddict{u}}{1}{\Nd-1}^{\ddict{\mt{p}}} \end{bmatrix}\right) = (1+\dnp)(\dnx+\dnu).
\end{equation}
\begin{prop}[{Data-driven LPV state-feedback synthesis~\cite{Verhoek2022_DDLPVstatefb}}]\label{prop:sfbsynthesis}%
	Given a $\datasetss$ that satisfies \eqref{eq:rkcond}. {If there exist $\tilde\lyap\posdef0$, } 
	$\Xi$, ${F}_Q$, $\mc{F}$ and $\mc{Y}$, such that
	\setlength{\dashlinegap}{1 pt}
	\begingroup\allowdisplaybreaks
	\begin{subequations}\label{eq:synthesis_conditions}
	\begin{gather}
		\left[\begin{array}{c} * \\ \hdashline  * \end{array}\right]^\top \left[\begin{array}{c:c} \Xi & 0 \\ \hdashline  0 & -W \end{array}\right]	 \left[\begin{array}{c c} L_{1,1} & L_{1,2} \\ I & 0 \\ \hdashline  L_{2,1} & L_{2,2} \end{array}\right]\prec 0, \label{eq:synthesis_conditions:a}\\
		\left[\begin{array}{c} * \\ \hdashline * \end{array}\right]^\top \underbrace{\begin{bsmallmatrix} \Xi_{1,1} & \Xi_{1,2} \\ \Xi_{1,2}^\top & \Xi_{2,2} \end{bsmallmatrix}}_{\Xi} \left[\begin{array}{c} I \\ \hdashline \Delta_p \end{array}\right]   \succeq 0, \quad \Xi_{2,2} \prec 0, \label{eq:synthesis_conditions:b}\\
		\begin{bmatrix} \tilde\lyap & 0 & 0 \\ 0 & I_{n_\mr{p}}\otimes {\tilde\lyap} & 0 \\ Y_0 & \bar{Y}& 0 \\ 0 & I_{n_\mr{p}}\otimes Y_0 &  I_{n_\mr{p}}\otimes \bar{Y} \end{bmatrix} = \begin{bmatrix} \mc{H}_1(\sigint{\ddict{x}}{1}{\Nd-1} \\ \mc{H}_1(\sigint{\ddict{x}}{1}{\Nd-1}^{\ddict{\mt{p}}} \\ \mc{H}_1(\sigint{\ddict{u}}{1}{\Nd-1} \\ \mc{H}_1(\sigint{\ddict{u}}{1}{\Nd-1}^{\ddict{\mt{p}}} \end{bmatrix}
		\mc{F}, \label{eq:synthesis_conditions:c}\\
		\begin{bsmallmatrix} I_{\Nd} \\ \msf{p}\kron I_{\Nd} \end{bsmallmatrix}^\top %
		{F_Q} \begin{bsmallmatrix} I_{\dnx} \\ \msf{p}\kron I_{\dnx}  \end{bsmallmatrix} = %
		{\mc{F}}\begin{bsmallmatrix} I_{\dnx} \\ \msf{p}\kron I_{\dnx} \\ \msf{p}\kron \msf{p}\kron I_{\dnx} \end{bsmallmatrix}, \label{eq:synthesis_conditions:d}
	\end{gather}
	\end{subequations}
	are satisfied for all $\msf{p}\in\mb{P}$, where
	\begin{subequations}\label{eq:thm:lqr_vars}
	\begin{align}
		\mc{Y} & =[\,Y_0 \ \ \bar{Y}\,], \quad {Y_0\in\mb{R}^{\dnu\times\dnx}, \ \bar{Y}\in\mb{R}^{\dnu\times\dnp\dnx},}\\
		\Delta_p & = \mr{blkdiag}\big( \msf{p}_{1}I_{2\dnx},\dots,\msf{p}_{n_\mr{p}}I_{2\dnx}\big), \\
		L_{1,1} & = 0_{2n_\mr{x}n_\mr{p}\times 2n_\mr{x}n_\mr{p}},\\
		L_{1,2} & = \begin{bmatrix} {1}_{n_\mr{p}}\otimes I_{2n_\mr{x}} & 0_{2n_\mr{x}n_\mr{p}\times (n_\mr{x}+n_\mr{u})} \end{bmatrix},\\
		L_{2,1} & = \begin{bmatrix} 0_{n_\mr{x}\times 2n_\mr{x}n_\mr{p}}\\ I_{n_\mr{p}}\otimes \Gamma_1\\ 0_{n_\mr{x}\times 2n_\mr{x}n_\mr{p}}\\ I_{n_\mr{p}}\otimes \Gamma_2\\ 0_{(n_\mr{x}+n_\mr{u})\times 2n_\mr{x}n_\mr{p}} \end{bmatrix}, \quad \begin{matrix} \Gamma_1 = [ \, I_{\dnx} \ \ 0 \,], \vphantom{\Big)} \\ \Gamma_2 = [\, 0 \ \ I_{\dnx}\,], \vphantom{\Big)}
		\end{matrix}\\
		L_{2,2} & = \begin{bmatrix} \Gamma_1 & 0 \\ {1}_{n_\mr{p}}\otimes 0_{n_\mr{x}\times 2n_\mr{x}}&0\\ \Gamma_2 &0\\ 1_{n_\mr{p}}\otimes 0_{n_\mr{x}\times 2n_\mr{x}}&0\\ 0 & I_{(n_\mr{x}+n_\mr{u})} \end{bmatrix}, \\
		W & = \begin{bmatrix} \tilde\lyap_0 & F_Q^\top \overrightarrow{\mc{X}}^\top  & \begin{bmatrix} \tilde\lyap Q^{\frac{1}{2}} \\ 0\end{bmatrix} & \mc{Y}^\top R^{\frac{1}{2}} \\ \overrightarrow{\mc{X}}F_Q & \tilde\lyap_0 & 0 & 0 \\ \begin{bmatrix}  Q^{\frac{1}{2}}\tilde\lyap  & 0\end{bmatrix} & 0 & I_{n_\mr{x}} & 0 \\ R^{\frac{1}{2}}\mc{Y} & 0 & 0 &  I_{n_\mr{u}} \end{bmatrix},
	\end{align}
	with $\tilde\lyap_0 = \mr{blkdiag}\big(\tilde\lyap, \ 0_{\dnx\dnp\times\dnx\dnp} \big)$, and $\overrightarrow{\mc{X}} = \mr{blkdiag}\big(\mc{H}_1(\sigint{\ddict{x}}{2}{\Nd}), \ I_{\dnp}\kron\mc{H}_1(\sigint{\ddict{x}}{2}{\Nd}) \big)$,
	\end{subequations}
	\endgroup
	{then with}
	\begin{equation}\label{eq:KSD-stab}
		K_0 = Y_0 {\tilde\lyap}^{-1}, \quad \begin{bmatrix} K_1\! &\! \cdots\! &\! K_{\dnp} \end{bmatrix} = \bar{Y} (I_{n_\mr{p}}\otimes {\tilde\lyap} )^{-1},
	\end{equation}
	{the LPV state-feedback controller \eqref{eq:fblaw} stabilizes $\Bss$. If $\tilde\lyap$ is such that {$\mr{trace}(\tilde\lyap)$} is minimal among all feasible solutions of \eqref{eq:synthesis_conditions}, then \eqref{eq:fblaw} minimizes the supremum of $J_\infty(x, u) = \sum_{k=0}^{\infty}\ell(u_k,x_k)$ along all solutions of the resulting closed-loop system.}
\end{prop}
{In Proposition~\ref{prop:sfbsynthesis}, conditions~\eqref{eq:synthesis_conditions:a},~\eqref{eq:synthesis_conditions:b} are the result of the full-block $S$-procedure, for which the variables $\Delta_p$ and $L_{i,j}$ are introduced. Accordingly, $L(\msf{p})^\top W L(\msf{p})$, with $L(\msf{p}) = L_{2,2}+L_{2,1}\Delta_{p}(I-L_{1,1}\Delta_{{p}})^{-1}L_{1,2}$, results in an LMI that is quadratic in $\msf{p}$, which one may recognize from the design of an LPV-LQR controller, with the decision variables $\tilde Z^{-1}$ and $\mc{Y}$ as the Lyapunov matrix and the transformed controller gains, respectively. The closed-loop LPV system is characterized in a fully data-driven way by $\overrightarrow{\mc{X}}F_Q$ and conditions~\eqref{eq:synthesis_conditions:c},~\eqref{eq:synthesis_conditions:d}, see~\cite{Verhoek2022_DDLPVstatefb} for further details {on handling noisy data and the conservatism of Proposition~\ref{prop:sfbsynthesis}}.}
{Hence, solving the LMI conditions~\eqref{eq:synthesis_conditions} yields, via~\eqref{eq:KSD-stab}, a terminal {data-driven} LPV state-feedback controller $K(p)$. Moreover, the controller}
 satisfies Condition~\ref{A:termingredients}.\ref{A:termingredients:qstab} {with} $V(x) = x^\top \big({\tilde\lyap}^{-1}\big) x$ as a Lyapunov function for the closed-loop system. Let $Z={\tilde\lyap}^{-1}$ for the remainder of this section.
We now present data-based methods for the computation of the terminal set, {by exploiting and {extending} Proposition~\ref{prop:sfbsynthesis}. That is, we present a method to \emph{jointly} synthesize a terminal controller, a $\termcost$, and an ellipsoidal $\termset$ with an extension of Proposition~\ref{prop:sfbsynthesis} (Section~\ref{sss:ext}), as well as a method to compute a polyhedral $\termset$ by exploiting the resulting {$K(p)$} coming from Proposition~\ref{prop:sfbsynthesis} (Section~\ref{sss:exp})}. For this we need the following definition:
\begin{definition}[PC-sets]
    A compact and convex set with a non-empty interior that contains the origin is called a \emph{PC-set}.
\end{definition}
Moreover, for the data-based computation of $\termset$, we assume throughout this section {that} the following {properties hold:} %
\begin{assumption}\label{A:polystateinpconset}
    The following {properties} are satisfied:
    \begin{enumerate}[label=\alph*), ref={\alph*}]
        \item 
            The scheduling set $\mb{P}\subset\R^{\dnp}$ is a convex polytope, generated by a finite number of vertices, i.e., 
            \( \mb{P} := \mr{co}\left( \{\msf{p}^\mr{v}_i\}_{i=1}^{n_\mr{v}} \right), \)
            where $\mr{co}$ denotes the convex hull.
        \item
            The state and input constraint sets $\mb{X}$ and $\mb{U}$ are polyhedrons that are described by
            \begin{subequations}\label{eq:poly-state-input-constraints}
            \begin{align}
                \mb{X} & := \{\msf{x} \in\R^{n_\mr{x}} \mid {g_i^\mr{x}}^\top \!\msf{x}\leq h^\mr{x}_i,\, i=1,\dots,n_\mr{gx}\},\\
                \mb{U} & := \{ \msf{u} \in\R^{n_\mr{u}} \mid {g_i^\mr{u}}^\top \!\msf{u}\leq h^\mr{u}_i,\, i=1,\dots,n_\mr{gu}\}, 
            \end{align}
            \end{subequations}
            where, $g^\mr{x}_i\in\R^{n_\mr{x}}$, $g^\mr{u}_i\in\R^{n_\mr{u}}$, and $h^\mr{x}_i,h^\mr{u}_i\in\R$.
    \end{enumerate}
\end{assumption}
This assumption allows to use computationally efficient \emph{linear programming} (LP) or \emph{quadratic programming} (QP)  for {constructing the terminal ingredients.} %
When we choose the terminal cost $\termcost$ as the Lyapunov function that guarantees stability of the closed-loop with $K(p)$, i.e., $V_\mr{f}({\predtr{x}_{\Nc|k}})=\predtr{x}_{\Nc|k}^\top \lyap \predtr{x}_{\Nc|k}$, we can guarantee exponential stability %
in case $\termset$ satisfies Condition~\ref{A:termingredients}. We will now give two methodologies for the computation of a \emph{maximum positively invariant} (MPI) set $\termset$, i.e., the computation of $\termset$ that maximizes the domain of attraction of the LPV-SS-DPC. The first method assumes an ellipsoidal $\termset$, while the second method assumes a polyhedral $\termset$. As in the model-based case, each method has its own advantages and {disadvantages} %
\cite{nguyen2014settheoretic}. %

\subsubsection{Joint computation of $\mathit{\termcost}$ and an ellipsoidal $\termset$}\label{sss:ext}
Given a $K(p)$ computed using Proposition~\ref{prop:sfbsynthesis} from which $\mc{Y}$ and $\lyap$ are obtained. A candidate ellipsoidal $\termset$ that satisfies Condition~\ref{A:termingredients} and Assumption~\ref{A:polystateinpconset} is a sub-level set of %
$\termcost$~\cite{rawlings2020model}, i.e.,
\[
    \Omega=\{\msf{x}\in\mb{X} \mid \msf{x}^\top \lyap \msf{x}\leq\alpha^2,\   K({\mathbb{P}})\msf{x} {\ \subseteq\ } \mb{U}\}.
\]
The scalar $\alpha$ is a parameter that is used to enlarge the set~$\Omega$. The MPI ellipsoidal $\termset$, i.e., the maximum $\Omega$, for a given~$\lyap$ and~{$K(p)$} can be obtained by maximizing the value of $\alpha^2$ subject to the constraints $\Omega\subseteq\mb{X}$ and $K(\msf{p})\Omega\in\mb{U}$ for all $\msf{p}\in\P$.
The maximization can be carried out jointly with the computation of 
$\lyap$ and the terminal controller {$K(p)$} under the maximization of the determinant or trace of $\lyap$, as the volume of the ellipsoid $\Omega$ is {determined} by the determinant or the trace of~$\lyap$. Hence, we merge the LPV state-feedback synthesis problem of Proposition~\ref{prop:sfbsynthesis} with the computation of an MPI ellipsoidal $\termset$. Using the Schur complement, the problem can be recasted in terms of the decision variable $\tilde{\lyap}\,(=Z^{-1})$. 
We consider here to maximize the logarithm of the determinant of $\tilde\lyap$ to preserve the convexity of the associated optimization problem~\cite{BoVa04}. Merging the two problems together with the incorporation of~\eqref{eq:poly-state-input-constraints}, results in the optimization problem:
\begin{subequations}\label{eq:MPI-ell-dd}
\begin{align}
    \max_{\tilde\lyap, \msf{p}\in\mb{P}}\quad & \log \det (\tilde\lyap)  \\
    \text{s.t. } \quad & \text{data-driven synthesis conditions \eqref{eq:synthesis_conditions}}, 
    \label{eq:MPI-ell-dd:b}\\
    & {g^\mr{x}_i}^\top \tilde\lyap g^\mr{x}_i \leq (h^\mr{x}_i)^2,\ \
    {{\forall}i\in\mathbb{I}_1^{n_\mr{gx}}},
     \label{eq:MPI-ell-dd:c}\\
    & \begin{bmatrix} 
        (h^\mr{u}_i)^2  &  {g^\mr{u}_i}^\top \mc{Y} \begin{bsmallmatrix} I_{n_\mr{x}} \\\msf{p}  \otimes I_{n_\mr{x}} \end{bsmallmatrix} \\
        (\ast)^\top & \tilde\lyap \end{bmatrix} \succeq 0, \ \ 
        {{\forall}i\in \mathbb{I}_1^{n_\mr{gu}}}.
        \label{eq:MPI-ell-dd:d}
\end{align}
\end{subequations}
Note that \eqref{eq:MPI-ell-dd:b} ensures invariance, %
whereas \eqref{eq:MPI-ell-dd:c} and~\eqref{eq:MPI-ell-dd:d} ensure
constraint satisfaction. 

We have now presented a fully data-based method for the computation of the terminal ingredients, with an ellipsoidal $\termset$, {which is in-fact rather simple to compute}. %
Note that in the model-based case, this is a common approach~\cite{nguyen2014settheoretic}. However, in that case, full model knowledge is required for the computation, and the input matrix $B$ must be %
scheduling \emph{independent}. Contrary, solving \eqref{eq:MPI-ell-dd} only requires a single sequence of data from an unknown LPV system, whose input matrix can be scheduling \emph{dependent}.
{The simplicity of the SDP~\eqref{eq:MPI-ell-dd} comes however at the cost of conservatism by restricting $\termset$ to be ellipsoidal.} For this reason, we also discuss the computation of the terminal ingredients with a more flexible terminal set in terms of a \emph{polyhedral} MPI. %

\subsubsection{Computation of a polyhedral $\termset$}\label{sss:exp}
The computation of a polyhedral MPI $\termset$ is based on the data-driven LPV representation of the closed-loop, i.e., the data-driven representation of $\phi_\mr{cl}$, 
\extver{see also~\cite[Thm.~1]{Verhoek2022_DDLPVstatefb}.}%
{see also Proposition~\ref{prop:closed-loop-data-based-general} in Appendix~\ref{appendix:sfb}.} 
Polyhedral invariant sets are more {favorable} for LP/QP-based predictive control problems, {as} they are generally more flexible than ellipsoidal sets, leading to a larger domain of attraction. This is, however, at the expense
of an increased representation complexity in terms of the number of constraints.
Furthermore, contrary to the computation of ellipsoidal MPI sets, computing a polyhedral MPI set is often carried out with LP tools~\cite{nguyen2014settheoretic}. This implies that the computation algorithm cannot handle quadratic scheduling dependency in $\phi_\mr{cl}(x,p)$. This can be avoided by designing a scheduling independent controller ${K(p) =K_0}$ {(see \cite[Rem.~1.ii]{Verhoek2022_DDLPVstatefb})}. Let us denote the data-driven representation of $\Bss$ in closed-loop with ${K_0}$ by $\phi_\mr{cl,r}(x,p)$.
Next, for computing the polyhedral MPI set, let the state and input constraints in \eqref{eq:poly-state-input-constraints} be rewritten in the compact form $G_\mr{x} \msf{x}\leq h_\mr{x}$, and $G_\mr{u} \msf{u}\leq h_\mr{u}$,
respectively, where $ G_\mr{x}$, $h_\mr{x}$, $G_\mr{u}$, $h_\mr{u}$ collect the respective vectors $g_i^\mr{x}$, $h_i^\mr{x}$, $g_i^\mr{u}$, $h_i^\mr{u}$. Therefore, the state constraint set of the {closed-loop} system {$x_{k+1}=\phi_\mr{cl,r}(x_k,p_k)$ ensuring invariance} is defined as 
\begin{equation}\label{eq:Omega_0}
    \Omega=\{ \msf{x} \in\R^{n_\mr{x}}\mid G \msf{x}\leq h\},
\end{equation}
where \(G=\begin{bmatrix} G_\mr{x}^\top & (G_\mr{u} {K_0})^\top\end{bmatrix}^\top\) and $h=\begin{bmatrix} h_\mr{x}^\top & h_\mr{u}^\top \end{bmatrix}^\top$. Inspired by~\cite{nguyen2014settheoretic}, we define now the data-driven extension of a \emph{one-step admissible pre-image set} for a given $\Omega$, which is a PC-set used for the computation of the polyhedral MPI~$\termset$:
\begin{equation*}
    \mr{pre}_1(\Omega) :=\{\msf{x} \in\R^{\dnx} \mid G \phi_\mr{cl,r}(\msf{x},\msf{p}^\mr{v}_i) \leq h, \,  {\forall i \in\mathbb{I}_1^{n_\mr{v}}} %
    \}.
\end{equation*}
{The} model-based computation of a polyhedral set for time-varying systems in~\cite{nguyen2014settheoretic} {is extended} to the data-driven LPV setting in Algorithm~\ref{label_pseudocode_Algorithm_DB-polyMPIset}, which computes a polyhedral MPI $\termset$ for the LPV-SS-DPC scheme using only $\datasetss$. 
\begin{algorithm}[t]
\SetKwInput{KwInput}{Input}                %
\DontPrintSemicolon
  \textbf{initialization: } Set ${\nu}=0$ and  $\Omega_0=\Omega$ as in \eqref{eq:Omega_0} \;
   \textbf{repeat}\;
   {
   	\quad	 $\Omega_{{\nu}+1} \gets \mr{pre}_1(\Omega_{\nu})\cap\Omega_{\nu}$\;
	\quad ${\nu}\gets {\nu}+1$\;
	}
 \textbf{until} $\Omega_{{\nu}}\supseteq\Omega_{\nu-1}$\;
 \textbf{return} $\mathbb{X}_\mr{f}=\Omega_{{\nu}}$ 
\caption{Data-based polyhedral MPI set}
\label{label_pseudocode_Algorithm_DB-polyMPIset}
\end{algorithm}
Note that each iteration of Algorithm~\ref{label_pseudocode_Algorithm_DB-polyMPIset} requires to remove redundant inequalities of $\Omega_{\nu+1}$ and subset testing, meaning that complexity of these calculations grows with the number of constraints of the obtained $\Omega_{\nu+1}$.

\section{Applying the LPV-DPC schemes}\label{s:practicalproblems}
{In this section\extver{\footnote{See the extended version of this paper (on arXiv) for a derivation of the recursive implementations of the LPV-DPC schemes.}}{}, we discuss how to effectively apply the proposed LPV-DPC schemes to nonlinear systems or systems with exogenous effects, together with the implementation of the methods under noisy measurements.}

\vspace{-2mm}
\subsection{External and internal scheduling scenarios}\label{ss:schedulingscenarios}

In the LPV framework, the scheduling signal $p$ is considered to be an \emph{independent} signal, a free variable of the system. %
This allows linearity of the underlying scheduling-dependent behavior, and this property has been key in the establishment of powerful convex control synthesis and analysis methods on which the LPV framework builds on.
 
In some applications of LPV systems, $p$ is indeed an independent signal, such as outside temperature, precipitation, or the effect of an other subsystem in terms of a reference signal or control input in an upper control layer. In these cases, the used LPV model provides an exact representation of the original system and the scheduling is called \emph{external}.   
 
However, in the majority of applied LPV control,  LPV descriptions are used as surrogate models of {an} underlying nonlinear system, where $p$ is actually a function $\psi$ of the output, state, and input signals associated with the system. We refer to $\psi$ as the \emph{scheduling map}. Although, in these cases, the models are sometimes warningly labeled to be \emph{quasi}-LPV, in fact, this \emph{internal} relation is intentionally neglected in the LPV representation, assuming $p$ to be independently varying in a set containing all possible trajectories of $\psi(x,u)$ that can occur during operation of the system. This results in the \emph{embedding} of the original nonlinear system in a linear behavior, where the assumed freedom of $p$ only introduces conservativeness of the representation, i.e., increased size of the solution set, and conservativeness of LPV analysis or synthesis approaches, e.g., upper bounding the true $\ell_2$-gain of the system. 
 
In this section, we provide methods to manage Assumption~\ref{A:futureschedulingknown} in the LPV-DPC design for the cases when $p$ is an external signal or when $p$ is dependent on internal signals. We want to highlight that the \emph{gain-scheduling} approach is applicable for both cases, i.e., when $p$ is taken to be a constant over the prediction horizon of the LPV-DPC. %
We consider the cases where~$p$ is not known or measurable outside of the scope of this paper. %

\subsubsection{The external scheduling scenario}
In this scenario, the future of the scheduling $\sigint{p}{k}{k+\Nc{-1}}$ in the prediction horizon (denoted by $\psigint{\predtr{p}}{{0}}{{\Nc-1}}{k}$ in the derivations) is either \textit{(i)} exactly known upfront (in case of, e.g.,  the reference signal of the upper control layer), i.e., Assumption~\ref{A:futureschedulingknown} is trivially satisfied, or \textit{(ii)}~predicted for $\Nc$ steps in the future based on the current measured value $p_k$ and its past. 

There are many available scheduling prediction methods in the LPV-MPC literature. The majority of these methods can in fact be used in our LPV-DPC schemes. We give a short review of some of the available methods. 
In \cite{broomhead2014robust, satzger2017robust}, an offline-identified linear prediction model is employed to predict the scheduling, while in~\cite{morato2019novel} a recursive least-squares method is proposed. In \cite{abbas2022GPest}, Gaussian process regression is used to provide effective prediction of the future scheduling trajectory, which allows to characterize the uncertainty or prediction error in the scheduling as well. To take  uncertainty of the predicted\footnote{The method in \cite{hanema2018anticipative} can be also used when no prediction of the future trajectory is available, considering a growing tube around the last observed $p_k$ in the prediction interval, corresponding to the classical, but often rather conservative, setting of LPV-MPC control.} scheduling trajectories into account, in~\cite{hanema2018anticipative}, a tube-based robust LPV-MPC control method is proposed with formal stability and recursive feasibility guarantees. While the former approaches provide prediction methods for externally scheduled setting, the latter approach can be used to incorporate prediction error or uncertainty in our LPV-DPC schemes, which is the subject of future work.

\subsubsection{The internal scheduling scenario}\label{sss:internalsched}
\begin{figure}
    \centering
    \includegraphics[width=\linewidth]{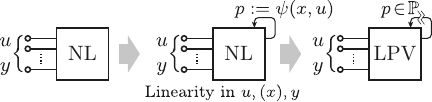}
    \caption{{Global LPV embedding of a nonlinear system with inputs $u$ and outputs $y$ and state realization $x$. A scheduling variable $p:=\psi(x,u)$ is defined such that if the trajectory of $p$ is known, then the remaining signal relations of {$y$ and $u$} are linear. To obtain an LPV representation, {the connection between $\psi(x,u)$ and $p$ is severed and} $p$ is assumed to be varying independently from $w$ in a bounded set $ \mb{P}  \supseteq \psi(\mathbb{X},\mathbb{U})$.}}  \vspace{-3mm}
    \label{fig:embedding}
\end{figure}
When LPV control is applied to a nonlinear system in the model-based case, an LPV surrogate model of the system is developed based on a \emph{global} LPV embedding of the NL behavior, depicted in Fig.~\ref{fig:embedding}, or \emph{local} LPV {modeling}  by means of interpolating local linearizations of the nonlinear system, see, e.g., \cite{Toth2010_book} for more details. For applying our DPC scheme, we consider global embedding  of a nonlinear system (Step 1 in Fig.~\ref{fig:embedding}) given by
\begin{equation}\label{eq:NLsys}
    x_{k+1} = f(x_k,u_k), \quad y_k = x_k.
\end{equation}
in the solution set of an LPV representation. The first
step is to write \eqref{eq:NLsys} as 
\begin{equation}\label{eq:NLsysrewritten}
    x_{k+1} = f_\mathrm{A}(x_k,u_k)x_k + f_\mathrm{B}(x_k,u_k)u_k, \quad y_k = x_k.
\end{equation}
{which can always be achieved if $f$ is continuously differentiable and $f(0,0)=0$, see \cite{Toth23CDCj}, where the latter is often ensured by appropriate state-transformation.} 
Then, a scheduling map $\psi:\mb{X}\times\mb{U}\to\mb{P}$ is defined that gives
\begin{equation}\label{eq:schedulingmap}
    p_k:=\psi(x_k,u_k),
\end{equation}
such that the resulting $A(p) = A(\psi(x,u)) = f_A(x,u)$ and $B(p) = B(\psi(x,u))= f_B(x,u)$ have a chosen class of functional dependency: affine, polynomial, rational, etc.
 This allows to write \eqref{eq:NLsysrewritten} as
\begin{equation}\label{eq:NLsyslpv}
    x_{k+1} = A(p_k)x_k + B(p_k)u_k.
\end{equation}
{corresponding to Step 2  in Fig.~\ref{fig:embedding}}.
Note that~\eqref{eq:NLsyslpv} and~\eqref{eq:NLsys} still exhibit the same behavior through~\eqref{eq:schedulingmap}. 
{To obtain an LPV representation, all allowed trajectories of~$p$ are restricted to a compact (often convex) set~$\mb{P}\supseteq \psi(\mb{X},\mb{U})$. Then, it is assumed that~$p$ varies \emph{independently} in the set~$\mb{P}$, which means that the actual relation between $\psi(x,u)$ and~$p$ in~\eqref{eq:schedulingmap} is disregarded.}
Due to the assumed independence of $(u,p,x)$, the behavior $\mf{B}_\mathrm{NL}$ associated with \eqref{eq:NLsyslpv} satisfies $\mf{B}_{\mathrm{NL}} \subseteq \pi_{(x,u)}\mf{B}$, corresponding to conservativeness of the representation as a price for linearity, but the original behavior $\mf{B}_\mathrm{NL}$  is always contained in $\mf{B}$.

Based on a given $\psi$, we will now provide a method to tackle Assumption~\ref{A:futureschedulingknown}, {i.e., obtain $\psigint{\predtr{p}}{{0}}{{\Nc-1}}{k}$ for the prediction horizon. A simple approach is known as \emph{gain-scheduled} predictive control, where $\psigint{\predtr{p}}{{0}}{{\Nc-1}}{k}$ is set to be equal with $p_k$ for all $k\in\mb{I}_0^{\Nc-1}$. While this method often works unexpectedly well in practice, it disregards variation of $A$ and $B$ through the $\psi$-represented connection with the optimization variables. An approach that is in line with the concept of the global embedding is to iteratively synthesize the scheduling trajectory}
as proposed in \cite{CiWe20}. %
The approach in~\cite{CiWe20} %
{calculates the scheduling trajectory over $\Nc$ based on the solution of $\predtr{x}$ and $\predtr{u}$ obtained in a previous iteration, iteratively synthesizing $(\predtr{u},\predtr{p},\predtr{x})$ through a sequence of QPs. The core idea of this scheme is extended to the data-driven setting in terms of Algorithm~\ref{alg:iterative-p-estim}.}
\begin{algorithm}[t]
\SetKwInput{KwInput}{Input}                %
\DontPrintSemicolon
  \textbf{initialization:} set $k\!=\!0$ and $\{\predtr{x}_{i+1|0}\!=\!x_0,\predtr{u}_{i|0}\!=\!0\}_{i=0}^{\Nc\unaryminus1}$\;
 \textbf{loop}\; 
 {
 \quad \textbf{repeat}\;  
 {
  \quad\quad   \textbf{update} $\predtr{p}_{[0,\Nc\unaryminus1]|k}$ by $\psi(\predtr{x}_{[0,\Nc\unaryminus1]|k},\predtr{u}_{[0,\Nc\unaryminus1]|k})$  \;
   \quad \quad  \textbf{solve} \eqref{eq:lpvdpc_opt:ss} to obtain $\predtr{x}^*_{[1,\Nc]|k}$, $\predtr{u}^*_{[0,\Nc\unaryminus1]|k}$ \;
    \quad\quad  \textbf{set} $\predtr{x}_{[1,\Nc\unaryminus1]|k}\gets\predtr{x}^*_{[1,\Nc\unaryminus1]|k}$ \;
    \quad\quad\hphantom{set }$\predtr{u}_{[0,\Nc\unaryminus1]|k}\gets \predtr{u}^*_{[0,\Nc\unaryminus1]|k}$ \;
    }
    \quad  \textbf{until}  $\predtr{u}_{[0,\Nc-1]|k}$ has converged\;
    \quad  \textbf{apply} $u_k=u_{0|k}$ and observe $x_{k+1}$\;
    \quad   \textbf{set} $k\gets k+1$ \;
   \quad   \textbf{set} $\predtr{x}_{[1,\Nc\unaryminus1]|k}\gets \predtr{x}_{[2,\Nc]|k-1}$ {and $\predtr{x}_{0|k}=x_k$}\;
   \quad  \hphantom{set }$\predtr{u}_{[0,\Nc\unaryminus2]|k}\gets \predtr{u}_{[1,\Nc-1]|k-1}$ and $\predtr{u}_{\Nc\unaryminus1|k}=0$. \;
   }
    \textbf{end loop}
\caption{LPV-SS-DPC for nonlinear systems}
\label{alg:iterative-p-estim}
\end{algorithm}
{By {seeing} the iterative procedure as inexact Newton steps for a root finding problem,~\cite{hespe2021convergence} shows under assumptions similar to those for standard SQP methods that the sequence of QPs has a local contraction property around a feasible, suboptimal solution of the LPV-DPC problem.} %

{Additionally, for the IO scheme under LPV embedding of a nonlinear system, the setpoint equilibrium $(u^\mr{r}, p^\mr{r}, y^\mr{r})$ can be determined based on only $y^\mr{r}$ by combining Remark~\ref{remark:artificialeq} and a simplified version of Algorithm~\ref{alg:iterative-p-estim}, as $p^\mr{r}$ is determined by $y^\mr{r}$ through $\psi$.}

\subsection{Working with noisy data}\label{ss:noise}

In practical applications of data-driven control, {measurement noise}
inevitably affects the behavior of closed-loop systems.
{Next}, we propose a \emph{robust} {modification of the} LPV-DPC scheme {{using} slack variables and regularization terms, which {makes it possible to} handle disturbances in the data-driven LPV representation}.
We {only discuss the modification of the} state-feedback case, but the same arguments are applicable to %
the output-feedback case.  %

Suppose that the system is affected by {a} bounded disturbance signal in an ARX-type setting\footnote{{Following the reasoning in~\cite{chiuso2023harnessing}, we can also handle other noise structures, e.g., Box-Jenkins, by increasing the order, i.e., $\tau$.}}, both in the offline {collected} data
$\sigint{\ddict{x}}{1}{\Nd}$ entering the Hankel matrices in %
Proposition~\ref{prop:FLeasy}, as well as in the {online measured} $x_k$ -- used as initial condition.
More precisely, we can only measure
\begin{align*}
    \sigint{\ddict{z}}{1}{\Nd}=\sigint{\ddict{x}}{1}{\Nd}+\sigint{\ddict{\varepsilon}}{1}{\Nd},\quad\text{and}\quad
    z_k=x_k+\varepsilon_k
\end{align*}
with a $\varepsilon_k$ with uniform distribution $\varepsilon_k \sim \mathcal{U}(-\varepsilon_\mr{max},\varepsilon_\mr{max})$ for all $k\in\mathbb{Z}$, which gives that
$\lVert\sigint{\ddict{\varepsilon}}{1}{\Nd}\rVert_{\infty}\leq\varepsilon_\mr{max}$ 
and $\lVert{\varepsilon}_k\rVert_{\infty}\leq\varepsilon_\mr{max}$. %
{We assume that the measurements of the scheduling~$\ddict{p}_k$ in~$\dataset$, as well as the online measurements~$\inittr{p}_k$, are noise-free.} 

In the literature, two main modifications have been proposed to cope with noise in LTI-DPC.
First, adding an additional slack variable to the data-dependent equality constraints in the prediction model~\eqref{eq:predictor:ss} was used %
for deriving closed-loop guarantees~\cite{berberich2020data}.
Inspired by this, we replace the nominal data-driven predictor from~\eqref{eq:predictor:ss} by the following robust one:
\begin{multline}\label{eq:predictor:ss:noise}
    \begin{bmatrix}
        \mc{H}_{1}(\sigint{\ddict{z}}{1}{\Nd-\Nc}) \\
        \mc{H}_{\Nc}(\sigint{\ddict{z}}{2}{\Nd}) \\
        \mc{H}_{\Nc}(\sigint{\ddict{u}}{1}{\Nd-1}) \\
        \mc{H}_{\Nc}(\sigint{\ddict{z}}{1}{\Nd-1}^{\ddict{\mt{p}}})-\mc{P}_k^{\dnx}\mc{H}_{\Nc}(\sigint{\ddict{z}}{1}{\Nd-1})\\
        \mc{H}_{\Nc}(\sigint{\ddict{u}}{1}{\Nd-1}^{\ddict{\mt{p}}})-\mc{P}_k^{\dnu}\mc{H}_{\Nc}(\sigint{\ddict{u}}{1}{\Nd-1}) 
    \end{bmatrix} g_k \\
    = \begin{bmatrix}
         z_k+\sigma_{0\mid k} \\
        \mr{vec}(\psigint{\predtr{z}}{1}{\Nc}{k}) +\sigma_{[1,\Nc]|k} \\
        \mr{vec}(\psigint{\predtr{u}}{0}{\Nc-1}{k}) \\ 
        \sigma_{[\Nc+1,2\Nc]|k} \\ %
        0%
    \end{bmatrix}.
\end{multline}
Here, {with a slight misuse of notation, the vector} $\sigma_{[0,2\Nc+1]|k}{\in\mb{R}^{\dnx(2\Nc+1)}}$  is a slack variable that is optimized online in order to relax the equality constraints and ensure feasibility.
    The main goal of adding the slack variable is to compensate the influence of the noise. %
To avoid an overly large prediction error due to the noise, the slack variable is regularized in the cost, i.e., we add a term $+\lambda_{\sigma}\lVert \sigma_{[0,2\Nc+1]|k}\rVert^2_2$.
In the LTI case, the regularization parameter $\lambda_{\sigma}>0$ is required to scale inversely with the noise level in order to prove practical stability, and we conjecture a similar connection in the LPV case.
The second modification is an additional regularization of $g_k$ that robustifies the DPC scheme against noise%
~\cite{markovsky2021behavioral}.
Thus, for the proposed robust LPV-DPC scheme, we add the term $\lambda_{\mathrm{g}}\lVert g_k\rVert^2_2$ to the cost.
Conversely to the regularization of the slack variable, the regularization parameter $\lambda_{\mathrm{g}}>0$ needs to scale directly (and not inversely) with the noise level to ensure theoretical guarantees.
In particular, for zero noise, the regularization is not required. {Note that the proposed regularization can be further refined. For example, it can be modified to the form of the $\Pi$-regularization~\cite{dorfler2022bridging}, which further reduces bias in the predictor}.

In summary, the following optimization problem defines the robust LPV-DPC scheme to control unknown LPV systems based on noisy data.
\begin{subequations}\label{eq:lpvdpc_opt:ss:noise}
    \begin{align}
        \min_{g_k,\sigma_{[0,2\Nc+1]|k}}  &  \termcost(\predtr{z}_{\Nc|k}-x^\mr{r})+{\textstyle\sum_{i=0}^{\Nc-1}}\ell(\predtr{u}_{i|k},\predtr{z}_{i|k}) 
        \nonumber \\
        &\quad +\lambda_{\mathrm{g}}\lVert g_k\rVert^2_2+\lambda_{\sigma}\lVert \sigma_{[0,2\Nc+1]|k}\rVert^2_2 \label{eq:lpvdpc_opt:ss:cost:noise}\\
        \text{s.t.} &\quad  \text{\eqref{eq:predictor:ss:noise} holds and } \predtr{z}_{0|k} = z_k, \\\label{eq:lpvdpc_opt:ss:input:noise}
        & \quad \predtr{u}_{i|k}\in\mb{U}, \ \ \forall i\in\mb{I}_0^{\Nc-1}, \ \ \predtr{z}_{\Nc|k}\in\termset.
    \end{align}
\end{subequations}
Note that problem~\eqref{eq:lpvdpc_opt:ss:noise} does not contain state constraints, which would require an additional constraint tightening due to the uncertain predictions (cf.~\cite{berberich2020constraints} for an output constraint tightening in robust LTI-DPC). {Furthermore,~\cite{Verhoek2022_DDLPVstatefb} also presents an extension of Proposition~\ref{prop:sfbsynthesis} with noisy data, allowing for data-driven computation of the terminal ingredients with noisy data.}
In the LTI case with noisy data, DPC with terminal cost and terminal region constraints can be shown to provide practical exponential stability guarantees (see~\cite{berberich2022inherent} for the main argument based on inherent robustness).
Here, practical exponential stability means that the closed-loop exponentially converges to a set around the setpoint, the size of which depends on the noise level~\cite{gruene2014asymptotic}.
Using an analogous approach to prove practical exponential stability of the proposed robust LPV-DPC scheme as well as handling noisy scheduling data are important objectives of future research.

\extver{}{
\subsection{Trading computational complexity via recursive LPV-DPC}
{In this section, we present two new LPV-DPC methods, which are closely related to the previously presented LPV-DPC schemes. These new LPV-DPC methods are the recursive versions of the LPV-IO-DPC and LPV-SS-DPC schemes. The recursive alternatives have advantages in terms of computational efficiency~\cite{koehler2022state}, allowing for a well-informed trade-off for computational complexity in the design of LPV-DPCs. This trade-off lies in the balance between the number of decision variables and the size of the problem.}
When the control horizon $\Nc$ is large or when the signal dimensions of the unknown LPV system are substantial, the matrices in the LPV-DPC can grow significantly in size. This results in a large optimization problem, which can in practice cause memory issues or problems with the computation time during operation. %
Instead of solving the problem for the full length-$\Nc$ trajectory at once, we recursively apply a one-step-ahead predictor  to obtain the prediction over the control horizon. For the formulation of {the two new LPV-DPC methods}, we use the derivations in Appendix~\ref{app:recursiveioform} for the recursive LPV-IO-DPC scheme, and the derivations in Appendix~\ref{app:olssrep} for the recursive LPV-SS-DPC scheme. The recursive version of the LPV-IO-DPC scheme is given as:
\begin{subequations}\label{eq:lpvdpc_opt:iorecursive}
\begin{align}
    \min_{g_k}  \quad & {\textstyle\sum_{i=0}^{\Nc-1}}\,\ell(\predtr{u}_{i|k},\predtr{y}_{i|k}) \label{eq:lpvdpc_opt:iorecursive:cost}\\
    \text{s.t.} \quad & \predtr{y}_{i|k} = {\bf\Phi}_\mr{ini}(\datasetio, \psigint{\predtr{p}}{i-\tau}{i}{k})\begin{bmatrix} \mr{vec}(\psigint{\predtr{u}}{i-\tau}{i-1}{k}) \\ \mr{vec}(\psigint{\predtr{y}}{i-\tau}{i-1}{k}) \end{bmatrix} \notag\\
    & \qquad + {\bf\Phi}_\mr{u}(\datasetio, \psigint{\predtr{p}}{i-\tau}{i}{k}) \, \predtr{u}_{i|k}, \\
    & \predtr{u}_{i|k}\in\mb{U},\ \predtr{y}_{i|k}\in\mb{Y} \quad i\in\mb{I}_0^{\Nc-1},\\
    & (\predtr{u}, \predtr{p}, \predtr{y})_{[\unaryminus\tau, \unaryminus1]|k} = (\inittr{u}, \inittr{p}, \inittr{y})_{[k-\tau, \tau-1]}, \\
    & \!\!\begin{bmatrix} u^\mr{r}_{\tau} \\
    y^\mr{r}_{\tau} \end{bmatrix} =\begin{bmatrix} \mr{vec}(\psigint{\predtr{u}}{\Nc-\tau}{\Nc-1}{k}) \\
    \mr{vec}(\psigint{\predtr{y}}{\Nc-\tau}{\Nc-1}{k})\end{bmatrix}, \label{eq:lpvdpc_opt:iorecursive:term}
\end{align}
where $\tau\geq\LBf$. We present here the scheme for a one-step-ahead IO-predictor. However, as highlighted in Appendix~\ref{app:recursiveioform}, this can also be formulated for $n$-step-ahead IO-predictors, which divides the prediction horizon up in larger portions.
\end{subequations}
The recursive formulation for the LPV-SS-DPC case is as follows:
\begin{subequations}\label{eq:lpvdpc_opt:ss:recursive}
\begin{align}
    \min_{g_k}  &\quad  \termcost(\predtr{x}_{\Nc|k}-x^\mr{r})+{\textstyle\sum_{i=0}^{\Nc-1}}\ell(\predtr{u}_{i|k},\predtr{x}_{i|k})\\
    \text{s.t.} &\quad  \predtr{x}_{i+1|k} = \mc{H}_1(\sigint{\ddict{x}}{2}{\Nd}) \begin{bmatrix} \mc{H}_1(\sigint{\ddict{x}}{1}{\Nd-1} \\ \mc{H}_1(\sigint{\ddict{x}}{1}{\Nd-1}^{\ddict{\mt{p}}} \\ \mc{H}_1(\sigint{\ddict{u}}{1}{\Nd-1} \\ \mc{H}_1(\sigint{\ddict{u}}{1}{\Nd-1}^{\ddict{\mt{p}}} \end{bmatrix}^\dagger \begin{bmatrix} \predtr{x}_{i|k} \\ \predtr{p}_{i|k}\kron \predtr{x}_{i|k} \\ \predtr{u}_{i|k} \\ \predtr{p}_{i|k}\kron \predtr{u}_{i|k}
	\end{bmatrix}, \label{eq:lpvdpc_opt:ss:recursive:model}\\
    & \quad \predtr{u}_{i|k}\in\mb{U}, \  \predtr{x}_{i+1|k}\in\mb{X}, \quad  i\in\mb{I}_0^{\Nc-1},\\
    & \quad \predtr{x}_{\Nc|k}\in\termset, \text{and }\predtr{x}_{0|k}=\inittr{x}_k. 
\end{align}
\end{subequations}
Note that \eqref{eq:lpvdpc_opt:ss:recursive:model} is now the one-step-ahead predictor for this recursive scheme, resembling~\eqref{eq:open-loop-data-based} in Appendix~\ref{app:olssrep}.

For both these recursive schemes, the decision variables are $\psigint{\predtr{u}}{0}{\Nc-1}{k}$ and $\psigint{\predtr{x}}{1}{\Nc}{k}/\psigint{\predtr{y}}{0}{\Nc-1}{k}$, i.e., in $\mb{R}^{\Nc(\dnu+\dnx)}$ or $\mb{R}^{\Nc(\dnu+\dny)}$, instead of only $g_k\in\mb{R}^{\Nd-\Nc+1}$ in the `multi-step' LPV-DPC schemes, while the problems themselves are much smaller, as less data is required to formulate the data-driven predictors. This is where the trade-off is, which remains an engineering choice. {We refer to~\cite{koehler2022state} for an in-depth discussion of benefits and limitations of recursive vs.~multi-step (data-driven) predictive control schemes, which apply analogously for the above LPV-DPC schemes.}

}

\section{Example: Unbalanced disc system}\label{s:example}

\begin{figure*}
    \centering
    \begin{minipage}[t]{.65\textwidth}
        \includegraphics[scale=1]{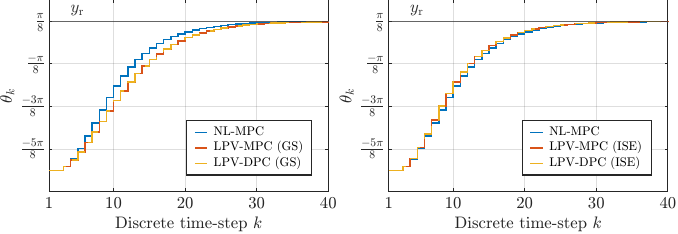}
        \caption{Simulation results for the comparison of predictive controllers that use IO measurements on the unbalanced disc system. We compare here a NL-MPC (\legendline{mblue}), an LPV-MPC (\legendline{morange}) and an LPV-DPC (\legendline{myellow}). The left plot shows the results where the LPV predictive controllers use GS to {determine the scheduling in the prediction horizon}, while the right plot shows the result where the LPV predictive controllers use ISE {for this purpose}. The simulation results show that our LPV-DPC method results in a similar performance as nonlinear and LPV model-based approaches.}
    \label{fig:resnoisefree}
    \end{minipage}\hfill
    \begin{minipage}[t]{.34\textwidth}
        \includegraphics[scale=1]{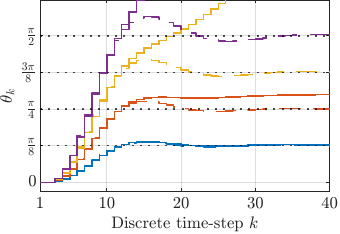}
        \caption{Comparison of the simulation results on the unbalanced disc system with a DeePC controller (\legendline{black}) and our LPV-IO-DPC controller \mbox{(\legendlined{black})} with different reference points. The further away $\theta_\mr{r}$ is from the stable equilibrium $\theta_k=0$, the more the DeePC performance degrades.}
    \label{fig:deepc}
    \end{minipage}
\end{figure*}

\subsection{Setup}
We demonstrate the applicability\footnote{{{Software implementation of the methods and the simulation environment are} available at: https://gitlab.com/releases-c-verhoek/lpvdpc}} of our results on a simulator of an unbalanced disc system. This system consists of a disc containing an off-centered mass, whose angle can be controlled by a DC motor. The nonlinear dynamics, discretized using {the} Euler method
with a sampling-time of $\ts=0.02$ seconds, are given by the state-space realization
\begin{subequations}\label{eq:unbalanceddisc}
\begin{align}
    \theta_{k+1} & = \theta_k + \ts\omega_k, \\
    \omega_{k+1} & = (1-\tfrac{\ts}{\tau_\mr{m}})\omega_k + \tfrac{\ts mgl}{J_\mr{m}}\sin(\theta_k) + \tfrac{\ts K_\mr{m}}{\tau_\mr{m}}u_{k},
\end{align}
\end{subequations}
where $\theta_k$ is the angular position in radians, $\omega_k$ is the angular speed in radians per second and $u_k$ the voltage in volts supplied to the DC motor. Moreover, $m,g,l,J_\mr{m}, K_\mr{m}$ and $\tau_\mr{m}$ are the model parameters\footnote{The model \eqref{eq:unbalanceddisc} of the system and parameter values taken from~\cite{abbas2021lpv} are only used for the model-based designs, data-generation, and validation of the controllers and not for the design of the data-driven controllers.}. Note that $\theta_k\equiv0$ corresponds to the upright position of the unbalanced disc, which is an unstable equilibrium. We can embed~\eqref{eq:unbalanceddisc} as an LPV-SS or LPV-IO representation by choosing\footnote{Note that the underlying (considered to be unknown) system is not an LPV system, but a nonlinear system, which requires that the map~$\psi$ in~\eqref{eq:schedulingmap} is assumed to be known.} $p_k=\sinc(\theta_k)$, from which we immediately see that $\mb{P}:=[-0.22,1]$. For this system, we will {compare our LPV-IO-DPC method with model-based IO methods and DeePC for noiseless and noisy data.} %
To determine the future scheduling trajectory in the prediction horizon we compare the \emph{gain-scheduling}~(GS) approach with the \emph{iterative scheduling estimation}~(ISE) approach via Algorithm~\ref{alg:iterative-p-estim}.  
Furthermore, 
we take~$\theta_k$ {to be the measured output of the system}. The {input and output constraints are defined as $u_k \in[-10, \, 10]=\mb{U}$ and $\theta_k \in [-\pi,\,\pi]=\mb{Y}$.}
The objective is to regulate the system to $(\theta_\mr{r}, u_\mr{r})=(\tfrac{\pi}{8}, -1.77)$, {where $u_\mr{r}$ is obtained as in Remark~\ref{remark:artificialeq}}. We first show the {comparison} on noise-free data, and finalize this example by demonstrating the performance of {the} LPV-IO-DPC with noise corrupted measurements. We want to emphasize that all the LPV-DPC designs discussed in this section are accomplished using only $\dataset$ and a set of tuning parameters. 

\subsection{Noise-free data} 
\vspace{-2pt}
\subsubsection{Comparison with model-based methods}\label{ss:examplenonoise}
{We first compare our LPV-IO-DPC methods to nonlinear IO-MPC (NL-MPC) and LPV-IO-MPC methods. For a fair comparison, we consider the same predictive control setup, except for the system representation. Hence, we use the same design parameters, and use the terminal equality constraints (i.e.,~\eqref{eq:lpvdpc_opt:io:term}) as terminal ingredients for all the schemes. The LPV schemes are solved with {\tt GUROBI}, while the NL-MPC is solved with {\tt IPOPT}. The design parameters are as follows.}
We choose $\Nc=20$, $\tau=2$ and the cost function is parametrized with $Q=R=1$. The terminal equality constraints are such that $(y_\mr{r}, u_\mr{r})=(\theta_\mr{r}, u_\mr{r})$. Because the underlying system is nonlinear, the scheduling in the LPV schemes is \emph{dependent} on~$y_\mr{r}$. To account for a possible mismatch in the prediction such that the terminal equality constraint yields the optimization problem infeasible, we have added a slack variable to the terminal output equality constraint, i.e., $\predtr{y}_{\Nc-i|k}=y_\mr{r}+\sigma_i$, $i\in\mb{I}_1^\tau$. The slack variable is penalized by adding $+10^7\|\sigma\|^2$ in the cost function. For the LPV schemes with the GS case, the scheduling signal is computed with the current measurement, i.e., $\predtr{p}_{i|k}=\sinc(y_k)$ for $i\in\mb{I}_0^{\Nc-1}$, while for the ISE case we follow Algorithm~\ref{alg:iterative-p-estim}.

The data-dictionary for the LPV-IO-DPC scheme is generated by applying $u_k\sim\mc{U}(\mb{U})$ to the system for $\Nd$ time steps. Based on Proposition~\ref{prop:FLeasy}, we have that for $\Nc=20$, $\tau=2$, $\Nd\geq89$, and thus chose $\Nd=89$. A posteriori verification of~\eqref{eq:thm:dim} gives that the obtained $\datasetio$ fully represents $\Bio|_{[1,\Nc+\tau]}$. 
Solving the predictive control problems %
yields the 
simulation results %
in Fig.~\ref{fig:resnoisefree}.

{From the simulation results, we can conclude that, for noise-free data, the LPV-MPC and LPV-DPC schemes are equivalent. This is expected because $\datasetio$ is able to fully represent the system behavior over the prediction horizon. However, one big advantage of the data-based scheme over  the model-based scheme is that the former only requires a data-set for the formulation of the predictive control scheme, while the latter requires an accurate model with exact knowledge of the model parameters $m,g,l,J_\mr{m},K_\mr{m},\tau_\mr{m}$. Furthermore, when compared to NL-MPC, which is solved as a nonlinear optimization problem,} {we see that the LPV-DPC achieves a similar performance. In fact, when the ISE method is used, the performance is even slightly better. The difference in} performance between the GS and ISE approach is because the ISE method {\emph{synthesizes} a scheduling sequence that corresponds to the planned solution of $y$ and $u$ in the prediction horizon.}

\subsubsection{Comparison with DeePC}
{We will now demonstrate the advantages of our methods over DeePC~\cite{coulson2019deepc} by means of a comparison on the unbalanced disc system with IO measurements. Note that DeePC has been introduced for LTI systems, hence when we tried to directly design a DeePC controller with~$\datasetio$ used in Section~\ref{ss:examplenonoise}, we did not manage to obtain a stabilizing DeePC controller. Therefore, we generated a new data-set with $\theta_k\equiv0$ as the downward vertical position, i.e., the stable equilibrium, and with $\ddict{u}_k\in[-2,2]$. This setting resulted in~$\breve\theta$ staying close to $0$. For the controller design, we added a regularization on $g_k$ with $\lambda = 0.01$ to the cost function and modified the stage cost to $\ell = (*)^\top Q(\predtr{y}_{i|k}-y^\mr{r}) + (*)^\top R(\predtr{u}_{i|k}-\predtr{u}_{i-1|k})$ with $Q=1$, $R=0.1$. For a fair comparison, we} {used the same stage cost in the LPV-IO-DPC controller (the extra regularization on $g$ was not included). The scheduling prediction over the control horizon, i.e., Assumption~\ref{A:futureschedulingknown}, is achieved with GS. We simulated the closed-loop system for the reference points $\theta_\mr{r}=\{\tfrac{\pi}{8},\tfrac{\pi}{4},\tfrac{3\pi}{8},\tfrac{\pi}{2}\}$, i.e., increasingly further away from the origin. The simulation results are depicted in Fig.~\ref{fig:deepc}, where the dotted lines represent the reference points, the dashed lines represent the simulations with the LPV-IO-DPC controller and the solid lines represent the simulations with the DeePC controller.
The plot shows that close to $\theta=0$, both controllers have a similar performance. Further away from the origin, however, the DeePC controller cannot handle the nonlinear effects of the unbalanced system behavior and even becomes unstable for $\theta_\mr{r}\geq \tfrac{3\pi}{8}$. On the other hand, the LPV-IO-DPC controller effortlessly regulates the nonlinear system to the {setpoint}, further emphasizing the advantages of using LPV-DPC over DeePC for data-driven control of nonlinear systems.}
\subsection{Handling noisy measurements}
We also demonstrate the performance with noisy measurements for a GS LPV-IO-DPC with the modifications discussed in Section~\ref{ss:noise} and compare it with the classic two-step approach, i.e., identification of an LPV-IO model followed by LPV-MPC design. For the data-generation, we {add a white}  noise signal $e_k\sim\mc{U}([-\varepsilon_\mr{max}, \varepsilon_\mr{max})$ {to the output} in an LPV-ARX setting. We choose~$\varepsilon_\mr{max}=0.01$, which corresponds to an angular error of at most $3.6$ degrees and a signal-to-noise ratio of $10$~dB, calculated w.r.t. the noise process. We again take $\Nd=89$.
Note that the noisy output signal also propagates through the model equations, corresponding to an ARX model structure. {Additionally, for illustration, we consider $p_k:=\mr{sinc}(y_k + e_k)$, i.e., a \emph{noisy} measurement of~$p_k$.}

For the two-step approach, we identify an LPV-ARX model with $\dna=\dnb=\tau$. Using the standard settings in \lpvcore\footnote{\lpvcore is an open-source \matlab toolbox%
, see \texttt{lpvcore.net}.}, we identify an LPV-ARX model based on $\datasetio$ ($\Nd=89$) with {\tt lpvarx}. The identified model is used as a {$N_\mathrm{c}$}-step-ahead predictor in a standard LPV-MPC scheme with the same terminal equality constraints.
We take $Q=10$ and $R=0.05$, while all other design parameters are the same as for the noise-free design. Additionally, for the LPV-IO-DPC design, we tuned the regularization parameters to $\lambda_\sigma=10^9$ and $\lambda_g=0.01$. %

We compare the two designs in two scenarios where we regulate the unbalanced disc to $(y_\mr{r},u_\mr{r})=(\tfrac{\pi}{8},-1.77)$. In the first scenario, we have next to the noisy $\datasetio$, also noisy online measurements of $y$. 
\begin{figure}
    \centering
    \includegraphics[scale=1]{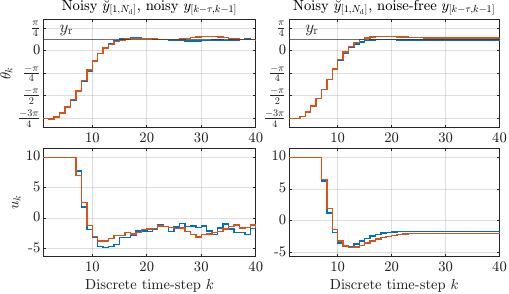}
    \caption{Comparison of the simulation results with the {indirect} two-step approach (\legendline{mblue}) and {the} LPV-IO-DPC (\legendline{morange}). For the left two plots, both $\ddict{y}$ and $\inittr{y}_{[k-\tau, k-1]}$ are affected by noise, while for the right two plots only $\ddict{y}$ is noisy and $\inittr{y}_{[k-\tau, k-1]}$ is measured noise free.}
    \label{fig:resnoise}
\end{figure}
The simulation results for this scenario are shown in the left two plots of Fig.~\ref{fig:resnoise}. To compare the performance of the two designs for handling a noisy data-dictionary, we have {a} second scenario, where the online measurements of $y$ are \emph{noise-free}. These results are shown in the right two plots of Fig.~\ref{fig:resnoise}. {On average, the added regularization terms and variables increased the computation time by 6 milliseconds, which corresponds to a 20\% increase compared to the noise-free case.}

The results show that both the two-step design and the LPV-IO-DPC design can regulate the unbalanced disc to a setpoint close to the upright position. What can be observed is that for the LPV-IO-DPC design, the regularization component seems to pull the input to zero. This phenomenon has been reported in multiple works on noise handling in LTI-DPC. Whether this effect can be negated by using more advanced regularization, e.g.,~\cite{markovsky2021behavioral}, for LPV-DPC requires further investigation.

\section{Conclusions%
}\label{s:conclusion}
In this work, we {have} derived novel output-feedback and state-feedback-based direct data-driven LPV predictive control schemes {using the} LPV Fundamental Lemma, which allows to {construct} a fully data-based predictor of {an} unknown LPV system. {M}ethods for computation of terminal ingredients {purely based on data are also} provided, {ensuring exponential stability and recursive feasibility of the closed-loop system regulated by the proposed DPC designs. Furthermore, effective approaches for handling noise and disturbances in the data are derived, together with methods for determining the scheduling sequence in case the approach is applied for nonlinear systems or systems dependent on exogenous effects. Through comparison of the developed DPC schemes with nonlinear, LPV and LTI MPC solutions and the LTI DeePC scheme on an unbalanced disc system, we have demonstrated that our proposed method, without the need of a modeling step, can achieve the same high performance as LPV and nonlinear predictive controllers which use an exact model of the system, while the LTI DeePC scheme was not able to stabilize the nonlinear system.}
As a future work, we aim to develop statistically efficient handling of noise in the proposed LPV DPC schemes and, via the concepts {detailed} in~\cite{Verhoek2023_DDCGenNLSys}, provide reference tracking under global stability and performance guarantees for nonlinear systems.

\appendix
\subsection{Proof of Lemma~\ref{lemma:detect}}\label{appendix:ioss}
For notational brevity, we denote {$\|\centerdot\|_2$ by $\|\centerdot\|$},  $p_k$ by $p$, and $p_{k+1}$ by $p^+$ throughout this proof, similarly for {$x$}. %
\begin{proof}
\begingroup\allowdisplaybreaks
As we consider behaviors for which there exists a state-minimal LPV-SS representation, we know that the corresponding behavior is controllable~\cite{petreczky2023minimal}. %
    Due to minimality and affine scheduling dependence, the LPV-SS realization is detectable, in fact, it is {completely state-observable} \cite{Toth2010_book, petreczky2023minimal}. This implies that there exists a parameter-varying gain matrix $L:\mb{P}\to\R^{\dnx\times\dny}$ and a positive definite $P:\mb{P}\to\R^{\dnx\times\dnx}$, i.e., $P({p})\posdef0$ for all ${p}\in\mb{P}$, such that 
    \begin{equation}\label{eq:lmdetect:I}
       P(p) - (A(p)+L(p)C(p))^\top P(p^+)(A(p)+L(p)C(p)) = I,
    \end{equation}
    for all $p,p^+\in\mb{P}$. {Then,} with the candidate IOSS Lyapunov function $V(x,p) = x^\top P(p)x$, {we} have the analog of~\cite[Eq. (12)]{cai2008input}:
    \begin{align}
        V(&x^+,p^+) - V(x,p) = (x^+)^\top P(p^+)x^+ - x^\top P(p)x \label{eq:lmdetect:lyap}\\
         =\,& (*)^\top P(p^+)((A(p)+L(p)C(p))x+B(p)u-L(p)y) \notag\\ 
         & \ - x^\top P(p)x \\
         =\,& x^\top(A(p)+L(p)C(p))^\top P(p^+)(A(p)+L(p)C(p))x \notag\\
         & \ + u^\top B^\top(p) P(p^+) B(p) u + y^\top L^\top (p)P(p^+) L(p) y\notag\\
         & \ - x^\top P(p)x + 2 x^\top (A(p)+L(p)C(p))^\top P(p^+)B(p)u \notag\\
         & \ - 2 x^\top (A(p)+L(p)C(p))^\top P(p^+)L(p) y \notag\\
         & \ -2u^\top B^\top(p) P(p^+)L(p) y \\
         \leq \, &\|u\|^2\|B(p)\|^2\|P(p^+)\|+ \|y\|^2\|L(p)\|^2\|P(p^+)\| \notag\\
         & \ -\|x\|^2 + 2\|u\|\|B(p)\|\|P(p^+)\|\|L(p)\|\|y\|\notag\\
         & \ + 2 \|x\|\|A(p)+L(p)C(p)\|\|P(p^+)\|\|B(p)\|\|u\| \notag \\ 
         & \ + 2 \|x\|\|A(p)+L(p)C(p)\|\|P(p^+)\|\|L(p)\|\|y\|. \label{eq:lmdetect:ineq}
    \end{align}
    We can upper bound the left-hand side of the inequality using~\eqref{eq:lmdetect:I} and $2ab\leq\tfrac{1}{4}a^2+4b^2$ by defining the constants:
    \begin{align*}
        \mu_A & = \max_{p\in\mb{P}}\|A(p)+L(p)C(p)\|, & \mu_B & = \max_{p\in\mb{P}}\|B(p)\|,  \\ 
        \mu_P & = \max_{p\in\mb{P}}\|P(p)\| = \max_{p^+\in\mb{P}}\|P(p^+)\|, & \mu_L & = \max_{p\in\mb{P}}\|L(p)\|,
    \end{align*}
    such that an upper bound of \eqref{eq:lmdetect:ineq} and thus \eqref{eq:lmdetect:lyap} is {of} the form:
    \begin{align}
        \eqref{eq:lmdetect:lyap}&\leq\eqref{eq:lmdetect:ineq}\leq \mu_B^2\mu_P\|u\|^2 + \mu_L^2\mu_P\|y\|^2 \notag\\
        &\hphantom{\leq\eqref{eq:lmdetect:ineq}\leq}+ 2\mu_A\mu_B\mu_P\|x\|\|u\|+ 2\mu_A\mu_L\mu_P\|x\|\|y\| \notag\\
         &\hphantom{\leq\eqref{eq:lmdetect:ineq}\leq}- \|x\|^2 + 2\mu_B\mu_P\mu_L\|u\|\|y\| \notag\\
         &\hphantom{\leq\eqref{eq:lmdetect:ineq}}\leq -\tfrac{1}{2}\|x\|^2 + (4\mu_A^2\mu_P^2 +\tfrac{5}{4}\mu_P )\mu_B^2\|u\|^2 \notag\\
         & \hphantom{\leq\eqref{eq:lmdetect:ineq}\leq}+ (4\mu_A^2\mu_P^2 +5\mu_P )\mu_L^2\|y\|^2 \notag\\
         & \hphantom{\leq\eqref{eq:lmdetect:ineq}} = -\tfrac{1}{2}\|x\|^2 + c_1\|u\|^2+ c_2\|y\|^2, \label{eq:lmdetect:const}
    \end{align}
    which is the upper bound in \eqref{eq:lmdetect:bound}.
    \endgroup
\end{proof}

\subsection{Proof of Theorem~\ref{thm:lpvssdpc}}\label{appendix:pf:thmlpvssdpc}
\begin{proof}
For proving recursive feasibility (i), we define a candidate input for the optimization problem~\eqref{eq:lpvdpc_opt:ss} by shifting the previously optimal solution
and appending {it by} the terminal controller from Condition~\ref{A:termingredients}:
$\predtr{u}_{i|k+1}=\predtr{u}^\ast_{i+1|k}$ {with $i\in\mathbb{I}_{0}^{\Nc-2}$} %
and $\predtr{u}_{\Nc-1|k+1}=K(\predtr{p}_{\Nc-1|k})\predtr{x}^\ast_{\Nc|k}$.  
By Assumption~\ref{alg:lpvdpc:ss}, this input and the corresponding state trajectory 
$\predtr{x}_{i|k+1}$ {with $i\in\mathbb{I}_{1}^{\Nc}$} %
satisfy constraints~\eqref{eq:lpvdpc_opt:ss:input}--\eqref{eq:lpvdpc_opt:ss:term}.
Further{more}, by Proposition~\ref{prop:FLeasy} {and the PE assumption}, there exists $g_{k+1}$ satisfying~\eqref{eq:predictor:ss} such that all constraints of Problem~\eqref{eq:lpvdpc_opt:ss} are satisfied.
Constraint satisfaction (ii) follows trivially from recursive feasibility.

Finally, to prove {exponential} stability (iii), we use the above {considered} candidate solution to arrive at
\begin{multline*}
    J_{\Nc}^*(k+1)-J_{\Nc}^*(k) \leq -\ell(\predtr{u}^\ast_{0|k},\predtr{x}_{0|k})+\termcost(\predtr{x}_{\Nc|k+1}) \\
    + \ell(K(\predtr{p}_{\Nc|k})\predtr{x}^\ast_{\Nc|k},\predtr{x}^\ast_{\Nc|k}) - \termcost(\predtr{x}^\ast_{\Nc|k}),
\end{multline*}
where $J_{\Nc}^*(k)\coloneqq J_{\Nc}^*(x_k,p_k,\psigint{\predtr{p}}{1}{\Nc}{k})$.
Using Condition~\ref{A:termingredients}, we obtain
\(%
    J_{\Nc}^*(k+1)-J_{\Nc}^*(k)
    \leq
    -\ell(\predtr{u}^\ast_{0|k},\predtr{x}_{0|k}).
\) %
Further{more}, {the following}  lower bound {trivially holds:}
\( %
    J_{\Nc}^*(k)\geq\eigmin(Q)\lVert\predtr{x}_{0|k}\rVert^2,
\) %
for any $\predtr{x}_{0|k}\in\mathbb{R}^{\dnx}$.
Finally, for any $x\in\mathbb{X}_{\mathrm{f}}$, the following upper bound holds
\( %
    J_{\Nc}^*(k)\leq \termcost(\predtr{x}_{0|k})\leq\eigmax(Z)\lVert x\rVert^2.
\) %
This can be extended to a quadratic upper bound on the set of all feasible initial states,  {see}~\cite[Prop.~2.16]{rawlings2020model}.
In conclusion, using standard Lyapunov arguments, the origin is exponentially stable for the closed-loop system. 
\end{proof}

\extver{}{
\subsection{Recursive LPV-IO formulation}\label{app:recursiveioform}
In this appendix, we formulate a data-driven recursive formulation of the LPV-IO representation. Consider the predictor for the LPV-IO-DPC scheme for $\Nc=1$, i.e., \eqref{eq:predictor:io}. We known from, e.g.,~\cite{MarkovskyRapisarda2008}, that the predicted output $y$ is completely determined for a given initial trajectory and a $u,p$ trajectory that is associated with $y$. That is, for a given $g_k$ that is the solution to
\begin{equation*}%
\newcommand{\vspc}{\vphantom{\big)_p^d}}
    \begin{bmatrix}
        \vspc\mc{H}_{\tau}(\sigint{\ddict{u}}{1}{\Nd-\tau}) \\
        \vspc\mc{H}_{\tau}(\sigint{\ddict{y}}{1}{\Nd-\tau}) \\
        \vspc\mc{H}_{1}(\sigint{\ddict{u}}{\tau+1}{\Nd}) \\
        \vspc\mc{H}_{\tau}(\sigint{\ddict{u}}{1}{\Nd\unaryminus\tau}^{\ddict{\mt{p}}})\unaryminus\inittr{\mc{P}}_k^{\dnu}\mc{H}_{\tau}(\sigint{\ddict{u}}{1}{\Nd\unaryminus\tau}) \\ 
        \vspc\mc{H}_{\tau}(\sigint{\ddict{y}}{1}{\Nd\unaryminus\tau}^{\ddict{\mt{p}}})\unaryminus\inittr{\mc{P}}_k^{\dny}\mc{H}_{\tau}(\sigint{\ddict{y}}{1}{\Nd\unaryminus\tau}) \\
        \vspc\mc{H}_{1}(\sigint{\ddict{u}}{\tau+1}{\Nd}^{\ddict{\mt{p}}})\unaryminus\predtr{\mc{P}}_k^{\dnu}\mc{H}_{1}(\sigint{\ddict{u}}{\tau+1}{\Nd})  \\ 
        \vspc\mc{H}_{1}(\sigint{\ddict{y}}{\tau+1}{\Nd}^{\ddict{\mt{p}}})\unaryminus\predtr{\mc{P}}_k^{\dny}\mc{H}_{1}(\sigint{\ddict{y}}{\tau+1}{\Nd}) 
    \end{bmatrix} g_k = \begin{bmatrix}
        \vspc\mr{vec}(\sigint{\inittr{u}}{k-\tau}{k-1}) \\ 
        \vspc\mr{vec}(\sigint{\inittr{y}}{k-\tau}{k-1}) \\ 
        \vspc \predtr{u}_k \\
        \vspc 0 \\
        \vspc 0 \\  
        \vspc 0 \\ 
        \vspc 0
    \end{bmatrix}\!,
\end{equation*}
the output $\predtr{y}_k$ can be determined with $\predtr{y}_k = \mc{H}_{1}(\sigint{\ddict{y}}{\tau+1}{\Nd}) g_k$. %
Let us denote the left-hand side matrix by $\Phi(\sigint{p}{k-\tau}{k})$, such that a given solution for $g_k$ is
\[ g_k = \big(\Phi(\sigint{p}{k-\tau}{k})\big)^\dagger \begin{bmatrix} \mr{vec}(\sigint{\inittr{u}}{k-\tau}{k-1}) \\ \mr{vec}(\sigint{\inittr{y}}{k-\tau}{k-1}) \\ \predtr{u}_k \\ 0_{(\tau+1)(\dnp+1)(\dnu+\dny)\times1} \end{bmatrix}. \]
This allows us to compactly write up the recursive formulation for $\predtr{y}_k$, omitting the time-intervals for brevity:
\begin{multline}
    \predtr{y}_k = \mc{H}_{1}(\ddict{y}) g_k = \\
    \mc{H}_{1}(\ddict{y}) \Big( \big( \Phi^\top(p) \Phi(p) \big)^{-1} \begin{bmatrix} \mc{H}_\tau(\ddict{u}) \\ \mc{H}_\tau(\ddict{y}) \end{bmatrix}^\top \begin{bmatrix} \mr{vec}(\inittr{u}) \\ \mr{vec}(\inittr{y}) \end{bmatrix} \\ 
    + \big( \Phi^\top(p) \Phi(p) \big)^{-1} \mc{H}_1(\ddict{u})^\top \predtr{u}_k \Big),
\end{multline}
which, with 
\begin{align*}
    {\bf\Phi}_\mr{ini}(\datasetio, \sigint{p}{k-\tau}{k}) & = \mc{H}_{1}(\ddict{y}) \big( \Phi^\top(p) \Phi(p) \big)^{-1} \begin{bmatrix} \mc{H}_\tau(\ddict{u}) \\ \mc{H}_\tau(\ddict{y}) \end{bmatrix}^\top, \\
    {\bf\Phi}_\mr{u}(\datasetio, \sigint{p}{k-\tau}{k}) & = \mc{H}_{1}(\ddict{y}) \big( \Phi^\top(p) \Phi(p) \big)^{-1} \mc{H}_1(\ddict{u})^\top,
\end{align*}
allows to write $\predtr{y}_k$ as:
\begin{multline}
    \predtr{y}_k = {\bf\Phi}_\mr{ini}(\datasetio, \sigint{p}{k-\tau}{k})\begin{bmatrix} \mr{vec}(\sigint{u}{k-\tau}{k-1}) \\ \mr{vec}(\sigint{y}{k-\tau}{k-1}) \end{bmatrix} + \\ 
    + {\bf\Phi}_\mr{u}(\datasetio, \sigint{p}{k-\tau}{k}) \, \predtr{u}_k,
\end{multline}
which can be used to simulate a system trajectory in a recursive setting. Note that we can apply this technique not only for $\Nc=1$, but in-fact for any $\Nc$.

\subsection{Data-driven state-feedback representations}\label{appendix:sfb}
In this appendix, we recap the open- and closed-loop data-driven LPV state-feedback representations from~\cite{Verhoek2022_DDLPVstatefb}, which we use for the terminal ingredients computation and the formulation of the recursive LPV-SS-DPC scheme.
\subsubsection{Open-loop representation}\label{app:olssrep}
As detailed in \cite{Verhoek2022_DDLPVstatefb}, we can obtain a data-driven representation of $\Bss$ by separating the coefficient matrices in \eqref{eq:LPVSSdependency} from $x$, $p$ and $u$ in \eqref{eq:LPVSS:state}, i.e.,
\begin{equation}\label{eq:open-loop-model-based}
	x_{k+1} = \mc{A} \begin{bmatrix} x_k \\ p_k \kron x_k \end{bmatrix} + \mc{B}\begin{bmatrix} u_k \\ p_k\otimes u_k \end{bmatrix},
\end{equation}
with $\mc{A}=\begin{bmatrix} A_0  & \cdots & A_{\dnp} \end{bmatrix}$ and $\mc{B}=\begin{bmatrix} B_0  & \cdots & B_{\dnp} \end{bmatrix}$. Then, by the linearity of $\Bss$ along $p$, the following holds
\begin{equation}\label{eq:LPVSysIdent}
	\mc{H}_1(\sigint{\ddict{x}}{2}{\Nd})=\mc{A}
	\begin{bmatrix} \mc{H}_1(\sigint{\ddict{x}}{1}{\Nd-1} \\ \mc{H}_1(\sigint{\ddict{x}}{1}{\Nd-1}^{\ddict{\mt{p}}} \end{bmatrix}+
	\mc{B} \begin{bmatrix} \mc{H}_1(\sigint{\ddict{u}}{1}{\Nd-1} \\ \mc{H}_1(\sigint{\ddict{u}}{1}{\Nd-1}^{\ddict{\mt{p}}} \end{bmatrix}.
\end{equation}
Then $\Bss$, i.e., the behavior corresponding to \eqref{eq:LPVSS:state}, can be fully characterized in terms of $\datasetss$ via:
\begin{subequations}\label{eq:open-loop-data-based}
\begin{equation}
	x_{k+1} = \mc{H}_1(\sigint{\ddict{x}}{2}{\Nd}) \sigint{\mcG}{1}{\Nd-1}^\dagger \begin{bsmallmatrix} x_k \\ p_k\otimes x_k \\ u_k \\ p_k\otimes u_k
	\end{bsmallmatrix},
\end{equation}
where
\begin{equation}
    \sigint{\mcG}{1}{\Nd-1}:= \begin{bmatrix} \mc{H}_1(\sigint{\ddict{x}}{1}{\Nd-1} \\ \mc{H}_1(\sigint{\ddict{x}}{1}{\Nd-1}^{\ddict{\mt{p}}} \\ \mc{H}_1(\sigint{\ddict{u}}{1}{\Nd-1} \\ \mc{H}_1(\sigint{\ddict{u}}{1}{\Nd-1}^{\ddict{\mt{p}}} \end{bmatrix}. 
\end{equation}
\end{subequations}
The representation \eqref{eq:open-loop-data-based} is well-posed, i.e., $\datasetss$ contains enough information to represent the characterizations for $\Bss$, if the following condition, formulated in \cite{Verhoek2022_DDLPVstatefb}, is satisfied:
\begin{condition}[Persistency of Excitation]\label{cond:rank-cond}
    $\datasetss$ is persistently exciting w.r.t. $\Bss$ if $\sigint{\mcG}{1}{\Nd-1}$ has 	has full \emph{row} rank, i.e., $\rankdef{\sigint{\mcG}{1}{\Nd-1}}=(1+n_\mr{p})(n_\mr{x}+n_\mr{u})$.
\end{condition}
Based on this condition, we see that we need at least $\Nd \ge 1+(1+n_\mr{p})(n_\mr{x}+n_\mr{u})$ data points. This condition has been also observed in LPV subspace identification \cite{Cox2021}, where the matrices $\mc{A},\mc{B}$ are estimated based on $\sigint{\mcG}{1}{\Nd-1}$ and $\mc{H}_1(\sigint{\ddict{x}}{2}{\Nd})$ after an estimate of the state-sequence has been obtained. Note that \eqref{eq:open-loop-data-based} can be seen as a data-based 1-step-ahead predictor for the trajectories in $\Bss$.

\subsubsection{Closed-loop data-driven representation}
Under the control-law in \eqref{eq:fblaw}, the following result from \cite{Verhoek2022_DDLPVstatefb} provides a fully data-driven closed-loop representation:
\begin{appendixprop}\label{prop:closed-loop-data-based-general}
	Given $\datasetss$ for which Condition~\ref{cond:rank-cond} is satisfied. Furthermore, let $\sigint{\mcG}{1}{\Nd-1}$ be defined as in \eqref{eq:open-loop-data-based} under $\datasetss$. Then, the closed-loop system, i.e., $\Bss$ in closed-loop with $K(p_k)$ under the control-law \eqref{eq:fblaw}, is represented equivalently as
	\begin{equation}\label{eq:data-based-CLLPV-state-feedback-general}
		x_{k+1}=\mc{H}_1(\sigint{\ddict{x}}{2}{\Nd}) \mcV \begin{bmatrix}
		x_k\\p_k\otimes x_k \\ p_k\otimes p_k\otimes x_k
		\end{bmatrix},
	\end{equation} 
	where $\mcV\in\mathbb{R}^{\Nd-1 \times n_\mathrm{x}(1+n_\mathrm{p}+n_\mathrm{p}^2) }$ is any matrix that satisfies
	\begin{equation}\label{eq:consist-cond-general}
		\begin{bmatrix}
		I_{n_\mr{x}} & 0& 0\\
		0& I_{n_\mr{p}}\otimes I_{n_\mr{x}}& 0\\
		K_0 & \bar{K} & 0\\
		0 & I_{n_\mr{p}}\otimes K_0 &  I_{n_\mr{p}}\otimes \bar{K} 
		\end{bmatrix} = \sigint{\mcG}{1}{\Nd-1}\mcV,
	\end{equation}
	where $\bar{K} = \begin{bmatrix} K_1 & \cdots & K_{\dnp} \end{bmatrix}$. 
\end{appendixprop}
With this data-driven representation of the closed-loop, the synthesis algorithm of Proposition~\ref{prop:sfbsynthesis} is derived as presented in~\cite{Verhoek2022_DDLPVstatefb}. We want to highlight that the synthesis algorithm aims to find a stabilizing controller in the subspace of $\mc{V}$'s that satisfy \eqref{eq:consist-cond-general}. In the synthesis algorithm, this subspace is defined in terms of \eqref{eq:synthesis_conditions:c} and the decision variables $\mc{F}$ and $F_Q$, which are linked with $\mc{V}$ via the relationship
\begin{equation*}
	\mcV \begin{bsmallmatrix} I_{\dnx} \\ \msf{p}\kron I_{\dnx} \\ \msf{p}\kron \msf{p}\kron I_{\dnx} \end{bsmallmatrix} \tilde\lyap =  \begin{bsmallmatrix} I_{\Nd} \\ \msf{p}\kron I_{\Nd} \end{bsmallmatrix}^\top F_Q \begin{bsmallmatrix} I_{\dnx} \\ \msf{p}\kron I_{\dnx}  \end{bsmallmatrix} = \mc{F}\begin{bsmallmatrix} I_{\dnx} \\ \msf{p}\kron I_{\dnx} \\ \msf{p}\kron \msf{p}\kron I_{\dnx} \end{bsmallmatrix}.
\end{equation*}
}

\bibliographystyle{IEEEtran}
\bibliography{references_lpvdpc.bib}

\begin{thebibliography}{10}
\providecommand{\url}[1]{#1}
\csname url@samestyle\endcsname
\providecommand{\newblock}{\relax}
\providecommand{\bibinfo}[2]{#2}
\providecommand{\BIBentrySTDinterwordspacing}{\spaceskip=0pt\relax}
\providecommand{\BIBentryALTinterwordstretchfactor}{4}
\providecommand{\BIBentryALTinterwordspacing}{\spaceskip=\fontdimen2\font plus
\BIBentryALTinterwordstretchfactor\fontdimen3\font minus
  \fontdimen4\font\relax}
\providecommand{\BIBforeignlanguage}[2]{{%
\expandafter\ifx\csname l@#1\endcsname\relax
\typeout{** WARNING: IEEEtran.bst: No hyphenation pattern has been}%
\typeout{** loaded for the language `#1'. Using the pattern for}%
\typeout{** the default language instead.}%
\else
\language=\csname l@#1\endcsname
\fi
#2}}
\providecommand{\BIBdecl}{\relax}
\BIBdecl

\bibitem{Toth2010_book}
R.~T\'{o}th, \emph{Modeling and Identification of Linear Parameter-Varying
  Systems}.\hskip 1em plus 0.5em minus 0.4em\relax Springer, 2010.

\bibitem{rawlings2020model}
J.~B. Rawlings, D.~Q. Mayne, and M.~M. Diehl, \emph{Model Predictive Control:
  Theory, Computation, and Design}.\hskip 1em plus 0.5em minus 0.4em\relax Nob
  Hill Pub, 2020.

\bibitem{Toth21aIET}
J.~Hanema, R.~T\'{o}th, and M.~Lazar, ``Stabilizing non-linear model predictive
  control using linear parameter-varying embeddings and tubes,'' \emph{IET
  Control Theory \& Applications}, pp. 1--18, 2021.

\bibitem{Casavola2008a}
A.~Casavola, D.~Famularo, and G.~Franz{\`{e}}, ``{{A predictive control
  strategy for norm-bounded LPV discrete-time systems with bounded rates of
  parameter change}},'' \emph{International Journal of Robust and Nonlinear
  Control}, vol.~18, pp. 714--740, 2008.

\bibitem{CiWe20}
P.~Cisneros and H.~Werner, ``Nonlinear model predictive control for models in
  quasi-linear parameter varying form,'' \emph{International Journal of Robust
  \& Nonlinear Control}, vol.~30, no.~10, pp. 3945--3959, 2020.

\bibitem{morato2020lpvmpcsurvey}
M.~M. Morato, J.~E. Normey-Rico, and O.~Sename, ``Model predictive control
  design for linear parameter varying systems: A survey,'' \emph{Annual Reviews
  in Control}, vol.~49, pp. 64--80, 2020.

\bibitem{gidon2021data}
D.~Gidon, H.~S. Abbas, A.~D. Bonzanini, D.~B. Graves, J.~M. Velni, and
  A.~Mesbah, ``Data-driven {LPV} model predictive control of a cold atmospheric
  plasma jet for biomaterials processing,'' \emph{Control Engineering
  Practice}, vol. 109, p. 104725, 2021.

\bibitem{irdmousa2019data}
B.~K. Irdmousa, S.~Z. Rizvi, J.~M. Veini, J.~Nabert, and M.~Shahbakhti,
  ``Data-driven modeling and predictive control of combustion phasing for
  {RCCI} engines,'' in \emph{Proc. of the 2019 American Control Conference},
  2019, pp. 1617--1622.

\bibitem{luo2017data}
X.~Luo, ``Data-driven predictive control for continuous-time linear parameter
  varying systems with application to wind turbine,'' \emph{International
  Journal of Control, Automation \& Systems}, vol.~15, pp. 619--626, 2017.

\bibitem{cisneros2020data}
P.~S. Cisneros, A.~Datar, P.~G{\"o}ttsch, and H.~Werner, ``Data-driven
  quasi-{LPV} model predictive control using {Koopman} operator techniques,''
  in \emph{Proc. of the 21st IFAC World Congress}, 2020, pp. 6062--6068.

\bibitem{piga2017direct}
D.~Piga, S.~Formentin, and A.~Bemporad, ``Direct data-driven control of
  constrained systems,'' \emph{IEEE Transactions on Control Systems
  Technology}, vol.~26, no.~4, pp. 1422--1429, 2017.

\bibitem{formentin2016direct}
S.~Formentin, D.~Piga, R.~T{\'o}th, and S.~M. Savaresi, ``Direct learning of
  {LPV} controllers from data,'' \emph{Automatica}, vol.~65, pp. 98--110, 2016.

\bibitem{bao2023learning}
Y.~Bao, H.~S. Abbas, and J.~Mohammadpour~Velni, ``A learning-and scenario-based
  {MPC} design for nonlinear systems in {LPV} framework with safety and
  stability guarantees,'' \emph{International Journal of Control}, pp. 1--20,
  2023.

\bibitem{dorfler2023data}
F.~D{\"o}rfler, ``Data-driven control: Part two of two: Hot take: Why not go
  with models?'' \emph{IEEE Control Systems Magazine}, vol.~43, no.~6, pp.
  27--31, 2023.

\bibitem{WillemsRapisardaMarkovskyMoor2005}
J.~C. Willems, P.~Rapisarda, I.~Markovsky, and B.~L.~M. De~Moor, ``A note on
  persistency of excitation,'' \emph{Systems \& Control Letters}, vol.~54,
  no.~4, pp. 325--329, 2005.

\bibitem{yang2015data}
H.~Yang and S.~Li, ``A data-driven predictive controller design based on
  reduced {Hankel} matrix,'' in \emph{Proc. of the 2015 Asian Control
  Conference}, 2015, pp. 1--7.

\bibitem{coulson2019deepc}
J.~Coulson, J.~Lygeros, and F.~D{\"o}rfler, ``Data-enabled predictive control:
  in the shallows of the {DeePC},'' in \emph{Proc. of the 2019 European Control
  Conference}, 2019, pp. 307--312.

\bibitem{berberich2020data}
J.~Berberich, J.~K{\"o}hler, M.~A. M{\"u}ller, and F.~Allg{\"o}wer,
  ``Data-driven model predictive control with stability and robustness
  guarantees,'' \emph{IEEE Transactions on Automatic Control}, vol.~66, no.~4,
  pp. 1702--1717, 2020.

\bibitem{verheijen2021data}
M.~Lazar and P.~C.~N. Verheijen, ``Offset--free data--driven predictive
  control,'' in \emph{Proc. of the 61st Conference on Decision and Control},
  2022, pp. 1099--1104.

\bibitem{vanwingerden2022data}
J.-W. van Wingerden, S.~P. Mulders, R.~Dinkla, T.~Oomen, and M.~Verhaegen,
  ``Data-enabled predictive control with instrumental variables: the direct
  equivalence with subspace predictive control,'' in \emph{Proc. of the 61st
  IEEE Conference on Decision and Control}, 2022, pp. 2111--2116.

\bibitem{CSMarticleDorfler}
I.~Markovsky, L.~Huang, and F.~D{\"o}rfler, ``Data-driven control based on the
  behavioral approach: From theory to applications in power systems,''
  \emph{IEEE Control Systems Magazine}, vol.~43, no.~5, pp. 28--68, 2023.

\bibitem{li2022data}
R.~Li, J.~W. Simpson-Porco, and S.~L. Smith, ``Data-driven model predictive
  control for linear time-periodic systems,'' in \emph{Proc of the 61st IEEE
  Conference on Decision and Control}, 2022, pp. 3661--3668.

\bibitem{lian2021koopman}
Y.~Lian, R.~Wang, and C.~N. Jones, ``Koopman based data-driven predictive
  control,'' \emph{arXiv preprint arXiv:2102.05122}, 2021.

\bibitem{gholaminejad2023stable}
T.~Gholaminejad and A.~Khaki-Sedigh, ``Stable data-driven {Koopman} predictive
  control: Concentrated solar collector field case study,'' \emph{IET Control
  Theory \& Applications}, vol.~17, no.~9, pp. 1116--1131, 2023.

\bibitem{berberich2022linear}
J.~Berberich, J.~K{\"o}hler, M.~A. M{\"u}ller, and F.~Allg{\"o}wer, ``Linear
  tracking {MPC} for nonlinear systems -- {Part II}: The data-driven case,''
  \emph{IEEE Transactions on Automatic Control}, vol.~67, no.~9, pp.
  4406--4421, 2022.

\bibitem{VerhoekAbbasTothHaesaert2021}
C.~Verhoek, H.~S. Abbas, R.~T{\'o}th, and S.~Haesaert, ``Data-driven predictive
  control for linear parameter-varying systems,'' in \emph{Proc. of the 4th
  IFAC Workshop on Linear Parameter Varying Systems}, 2021, pp. 101--108.

\bibitem{toth2011state}
R.~T{\'o}th, H.~S. Abbas, and H.~Werner, ``On the state-space realization of
  {LPV} input-output models: Practical approaches,'' \emph{IEEE Transactions on
  Control Systems Technology}, vol.~20, no.~1, pp. 139--153, 2011.

\bibitem{VerhoekTothHaesaertKoch2021}
C.~Verhoek, R.~T{\'o}th, S.~Haesaert, and A.~Koch, ``Fundamental lemma for
  data-driven analysis of linear parameter-varying systems,'' in \emph{Proc. of
  the 60th IEEE Conference on Decision and Control}, 2021, pp. 5040--5046.

\bibitem{Verhoek2022_DDLPVstatefb}
C.~Verhoek, R.~T{\'o}th, and H.~S. Abbas, ``Direct data-driven state-feedback
  control of linear parameter-varying systems,'' \emph{International Journal of
  Robust and Nonlinear Control}, vol.~35, no.~16, pp. 6955--6977, 2025.

\bibitem{newpaper}
C.~Verhoek, I.~Markovsky, S.~Haesaert, and R.~T{\'o}th, ``A behavioral approach
  for {LPV} data-driven representations,'' \emph{IEEE Transactions on Automatic
  Control}, pp. 1--14, 2025.

\bibitem{cisneros2018constrained}
P.~S. Cisneros, A.~Sridharan, and H.~Werner, ``Constrained predictive control
  of a robotic manipulator using quasi-{LPV} representations,'' in \emph{Proc.
  of the 2nd IFAC Workshop on Linear Parameter-Varying Systems}, 2018, pp.
  118--123.

\bibitem{morato2019novel}
M.~M. Morato, J.~E. Normey-Rico, and O.~Sename, ``Novel {qLPV MPC} design with
  least-squares scheduling prediction,'' in \emph{Proc. of the 3rd IFAC
  Workshop on Linear Parameter-Varying Systems}, 2019, pp. 158--163.

\bibitem{markovsky2021behavioral}
I.~Markovsky and F.~D{\"o}rfler, ``Behavioral systems theory in data-driven
  analysis, signal processing, and control,'' \emph{Annual Reviews in Control},
  vol.~52, pp. 42--64, 2021.

\bibitem{MarkovskyRapisarda2008}
I.~Markovsky and P.~Rapisarda, ``Data-driven simulation and control,''
  \emph{International Journal of Control}, vol.~81, no.~12, pp. 1946--1959,
  2008.

\bibitem{lian2023adaptive}
Y.~Lian, J.~Shi, M.~Koch, and C.~N. Jones, ``Adaptive robust data-driven
  building control via bilevel reformulation: An experimental result,''
  \emph{IEEE Transactions on Control Systems Technology}, pp. 1--17, 2023.

\bibitem{bemporad2002explicit}
A.~Bemporad, M.~Morari, V.~Dua, and E.~N. Pistikopoulos, ``The explicit linear
  quadratic regulator for constrained systems,'' \emph{Automatica}, vol.~38,
  no.~1, pp. 3--20, 2002.

\bibitem{cai2008input}
C.~Cai and A.~R. Teel, ``Input--output-to-state stability for discrete-time
  systems,'' \emph{Automatica}, vol.~44, no.~2, pp. 326--336, 2008.

\bibitem{petreczky2023minimal}
M.~Petreczky, R.~T{\'o}th, and G.~Merc{\`e}re, ``Minimal realizations of
  input-output behaviors by {LPV} state-space representations with affine
  dependence,'' \emph{IEEE Control Systems Letters}, 2023.

\bibitem{limon2008mpc}
D.~Lim{\'o}n, I.~Alvarado, T.~Alamo, and E.~F. Camacho, ``{MPC} for tracking
  piecewise constant references for constrained linear systems,''
  \emph{Automatica}, vol.~44, no.~9, pp. 2382--2387, 2008.

\bibitem{berberich2020IFACchangingsetpoints}
J.~Berberich, J.~K{\"o}hler, M.~A. M{\"u}ller, and F.~Allg{\"o}wer,
  ``Data-driven tracking {MPC} for changing setpoints,''
  \emph{IFAC-PapersOnLine}, vol.~53, no.~2, pp. 6923--6930, 2020.

\bibitem{berberich2021terminal}
------, ``On the design of terminal ingredients for data-driven {MPC},'' in
  \emph{Proc. of the 7\tss{th} IFAC Conference on Nonlinear Model Predictive
  Control}, vol.~54, no.~6.\hskip 1em plus 0.5em minus 0.4em\relax Elsevier,
  2021, pp. 257--263.

\bibitem{spin2024unified}
L.~M. Spin, C.~Verhoek, W.~M.~H. Heemels, N.~van~de Wouw, and R.~T{\'o}th,
  ``Unified behavioral data-driven performance analysis: A generalized plant
  approach,'' in \emph{Proc. of the 2024 European Control Conference}, 2024,
  pp. 894--899.

\bibitem{MillerSznaier2022}
J.~Miller and M.~Sznaier, ``Data-driven gain scheduling control of linear
  parameter-varying systems using quadratic matrix inequalities,'' \emph{IEEE
  Control Systems Letters}, vol.~7, pp. 835--840, 2022.

\bibitem{verhoek2024decoupling}
C.~Verhoek, J.~Eising, F.~D{\"o}rfler, and R.~T{\'o}th, ``{Decoupling parameter
  variation from noise: Biquadratic Lyapunov forms in data-driven LPV
  control},'' in \emph{Proc. of the 63\tss{rd} IEEE Conference on Decision and
  Control}, 2024, pp. 6761--6766.

\bibitem{nguyen2014settheoretic}
H.-N. Nguyen, \emph{Set Theoretic Methods in Control}.\hskip 1em plus 0.5em
  minus 0.4em\relax Cham: Springer International Publishing, 2014, pp. 7--42.

\bibitem{BoVa04}
S.~Boyd and L.~Vandenberghe, \emph{Convex Optimization}.\hskip 1em plus 0.5em
  minus 0.4em\relax Cambridge University Press, 2004.

\bibitem{broomhead2014robust}
T.~Broomhead, C.~Manzie, L.~Eriksson, M.~Brear, and P.~Hield, ``A robust model
  predictive control framework for diesel generators,'' in \emph{Proc. of the
  19th IFAC World Congress}, 2014, pp. 11\,848--11\,853.

\bibitem{satzger2017robust}
C.~Satzger, R.~de~Castro, and A.~Knoblach, ``Robust linear parameter varying
  model predictive control and its application to wheel slip control,'' in
  \emph{Proc. 20th IFAC World Congress}, 2017, pp. 1514--1520.

\bibitem{abbas2022GPest}
A.~Elkamel, A.~Morsi, M.~Darwish, H.~S. Abbas, and M.~H. Amin, ``Model
  predictive control of linear parameter-varying systems using gaussian
  processes,'' in \emph{Proc. of the 26th International Conference on System
  Theory, Control and Computing}, 2022, pp. 452--457.

\bibitem{hanema2018anticipative}
J.~Hanema, ``Anticipative model predictive control for linear parameter-varying
  systems,'' Ph.D. dissertation, Technische Universiteit Eindhoven, 2018.

\bibitem{Toth23CDCj}
J.~H. Hoekstra, B.~Cseppento, G.~I. Beintema, M.~Schoukens, Z.~Koll{\'a}r, and
  R.~T{\'o}th, ``Computationally efficient predictive control based on {ANN}
  state-space models,'' in \emph{Proc. of the 62nd IEEE Conference on Decision
  and Control}, 2023, pp. 6330--6335.

\bibitem{hespe2021convergence}
C.~Hespe and H.~Werner, ``Convergence properties of fast quasi-{LPV} model
  predictive control,'' in \emph{Proc. of the 60th Conference on Decision and
  Control}, 2021, pp. 3869--3874.

\bibitem{chiuso2023harnessing}
A.~Chiuso, M.~Fabris, V.~Breschi, and S.~Formentin, ``Harnessing uncertainty
  for a separation principle in direct data-driven predictive control,''
  \emph{Automatica}, vol. 173, p. 112070, 2025.

\bibitem{dorfler2022bridging}
F.~D{\"o}rfler, J.~Coulson, and I.~Markovsky, ``Bridging direct and indirect
  data-driven control formulations via regularizations and relaxations,''
  \emph{IEEE Transactions on Automatic Control}, vol.~68, pp. 883--897, 2022.

\bibitem{berberich2020constraints}
J.~Berberich, J.~K{\"o}hler, M.~A. M{\"u}ller, and F.~Allg{\"o}wer, ``Robust
  constraint satisfaction in data-driven {MPC},'' in \emph{Proc. 59th IEEE
  Conference on Decision and Control}, 2020, pp. 1260--1267.

\bibitem{berberich2022inherent}
------, ``Stability in data-driven {MPC}: an inherent robustness perspective,''
  in \emph{Proc. 61st IEEE Conference on Decision and Control}, 2022, pp.
  1105--1110.

\bibitem{gruene2014asymptotic}
L.~Gr{\"u}ne and M.~Stieler, ``Asymptotic stability and transient optimality of
  economic {MPC} without terminal conditions,'' \emph{Journal of Process
  Control}, vol.~24, pp. 1187--1196, 2014.

\bibitem{koehler2022state}
J.~K{\"o}hler, K.~P. Wabersich, J.~Berberich, and M.~N. Zeilinger, ``State
  space models vs. multi-step predictors in predictive control: Are state space
  models complicating safe data-driven designs?'' in \emph{Proc. of the 61st st
  IEEE Conference on Decision and Control}, 2022, pp. 491--498.

\bibitem{abbas2021lpv}
H.~S. Abbas, R.~T{\'o}th, M.~Petreczky, N.~Meskin, J.~Mohammadpour~Velni, and
  P.~J.~W. Koelewijn, ``{LPV} modeling of nonlinear systems: A multi-path
  feedback linearization approach,'' \emph{International Journal of Robust \&
  Nonlinear Control}, vol.~31, no.~18, pp. 9436--9465, 2021.

\bibitem{Verhoek2023_DDCGenNLSys}
C.~Verhoek, P.~J.~W. Koelewijn, S.~Haesaert, and R.~T{\'o}th, ``Direct
  data-driven state-feedback control of general nonlinear systems,'' in
  \emph{Proc. 62\tss{nd} IEEE Conference on Decision and Control}, 2023, pp.
  3688--3693.

\bibitem{Cox2021}
P.~B. Cox and R.~T{\'o}th, ``Linear parameter-varying subspace identification:
  A unified framework,'' \emph{Automatica}, vol. 123, p. 109296, 2021.

\end{thebibliography}

\begin{IEEEbiography}[{\includegraphics[width=1in,height=1.25in,clip,keepaspectratio]{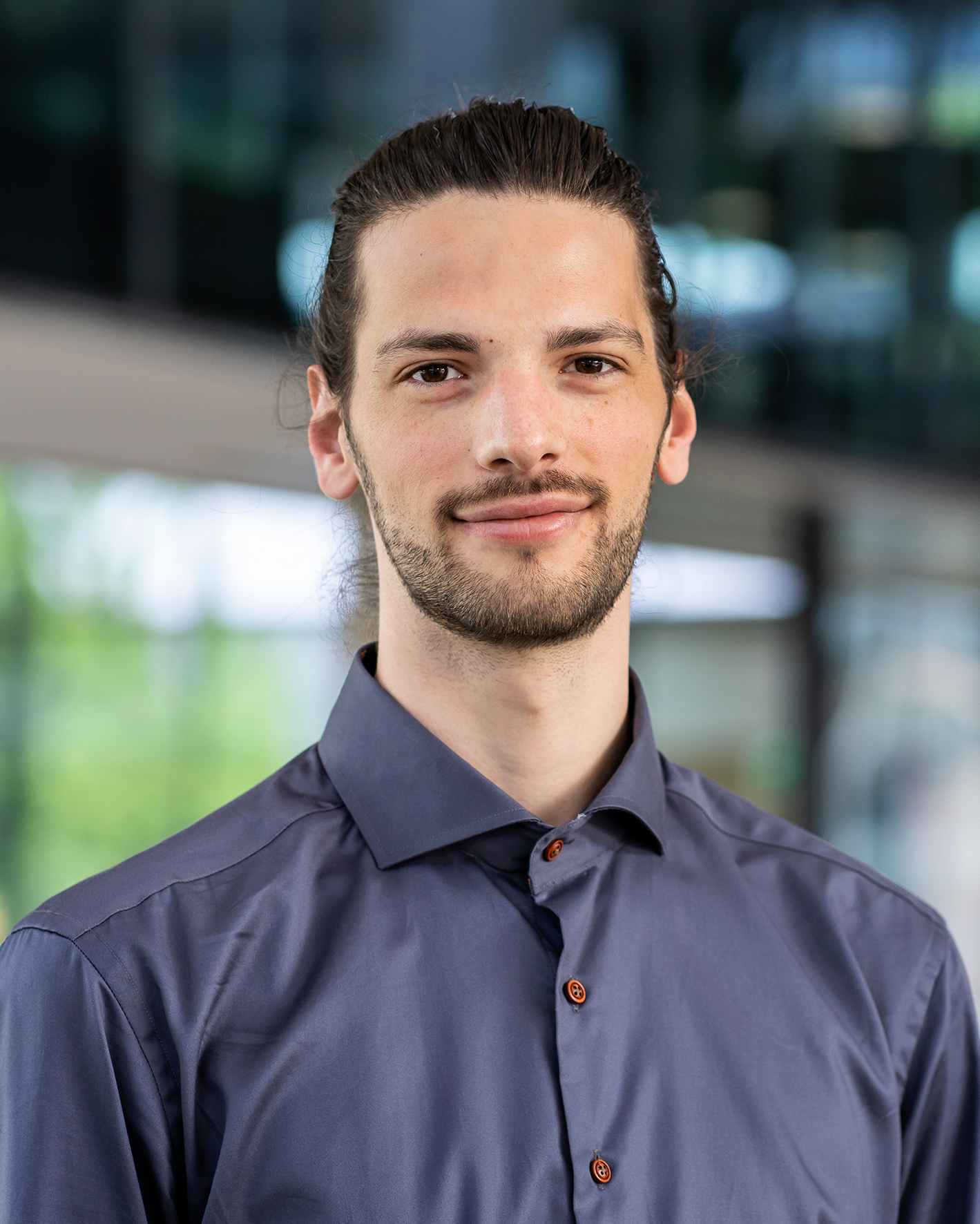}}]{Chris Verhoek} received his M.Sc. degree (Cum Laude) in Systems and Control from the Eindhoven University of Technology (TU/e) in 2020. In 2025, he received his Ph.D. degree with Cum Laude distinction, also from the TU/e. His M.Sc. thesis was selected as best thesis of the Electrical Engineering department in the year 2020. 
He is currently a postdoctoral researcher with the Control Systems group, Dept. of Electrical Engineering, TU/e. In the fall of 2023, he was a visiting researcher at the IfA, ETH Z{\"u}rich, Switzerland. His main research interests include (data-driven) analysis and control of nonlinear and LPV systems and learning-for-control techniques with stability and performance guarantees.
\end{IEEEbiography}
\begin{IEEEbiography}[{\includegraphics[width=1in,height=1.25in,clip,keepaspectratio]{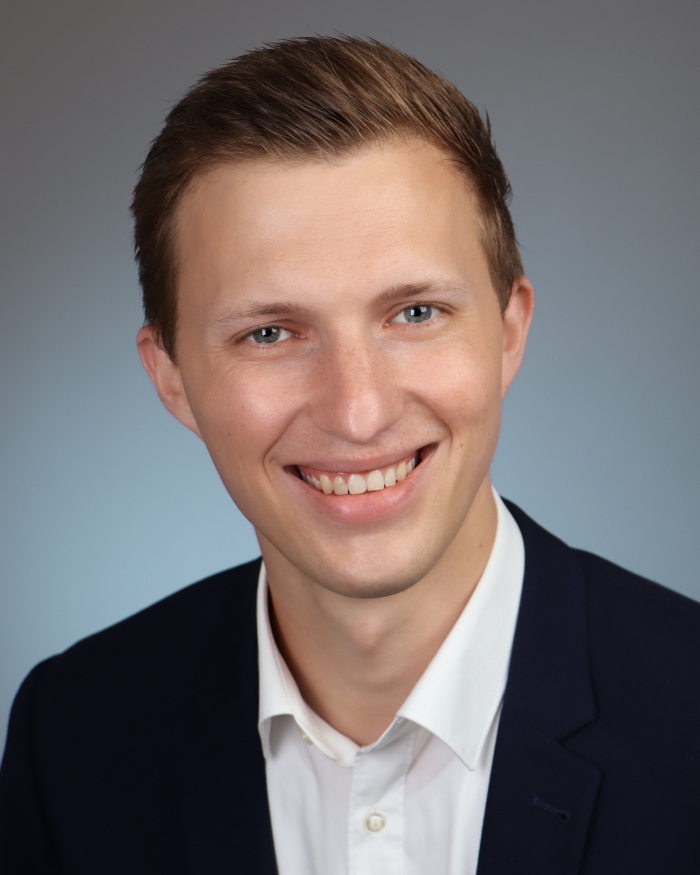}}]{Julian Berberich} received a Master's degree in Engineering Cybernetics from the University of Stuttgart, Germany, in 2018. In 2022, he obtained a Ph.D. in Mechanical Engineering, also from the University of Stuttgart, Germany. He is currently working as a Lecturer (Akademischer Rat) at the Institute for Systems Theory and Automatic Control at the University of Stuttgart, Germany. In 2022, he was a visiting researcher at the ETH Z{\"u}rich, Switzerland. He has received the Outstanding Student Paper Award at the 59th IEEE Conference on Decision and Control in 2020 and the 2022 George S. Axelby Outstanding Paper Award. His research interests include data-driven analysis and control as well as quantum computing.
\end{IEEEbiography}
\begin{IEEEbiography}[{\includegraphics[width=1in,height=1.25in,clip,keepaspectratio]{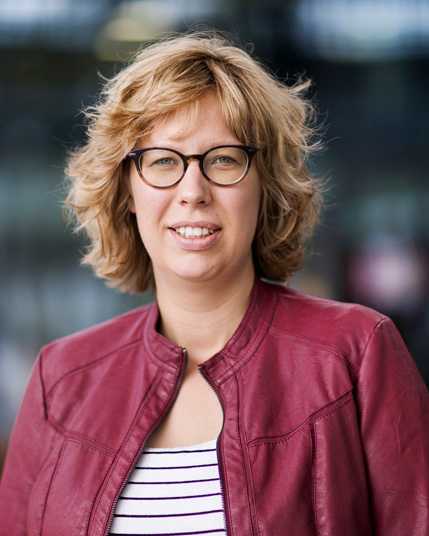}}]{Sofie Haesaert} received the B.Sc. degree cum laude in mechanical engineering and the M.Sc. degree cum laude in systems and control from the Delft University of Technology, Delft, The Netherlands, in 2010 and 2012, respectively, and the Ph.D. degree from Eindhoven University of Technology (TU/e), Eindhoven, The Netherlands, in 2017. She is currently an Assistant Professor with the Control Systems Group, Department of Electrical Engineering, TU/e. From 2017 to 2018, she was a Postdoctoral Scholar with Caltech. Her research interests are in the identification, verification, and control of cyber-physical systems for temporal logic specifications and performance objectives.
\end{IEEEbiography}
\begin{IEEEbiography}[{\includegraphics[width=1in,height=1.25in,clip,keepaspectratio]{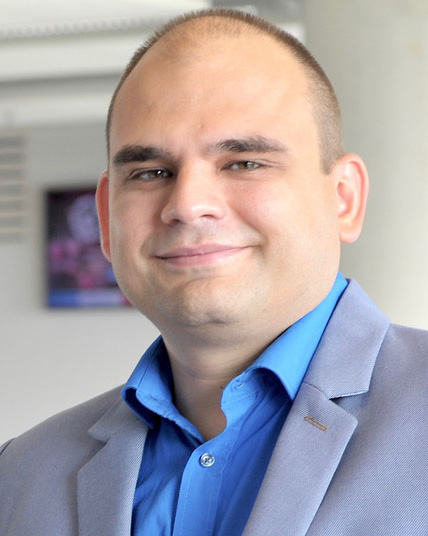}}]{Roland T\'oth} received his Ph.D. degree with Cum Laude distinction at the Delft University of Technology (TUDelft) in 2008. He was a post-doctoral researcher at TUDelft in 2009 and at the Berkeley Center for Control and Identification, University of California in 2010. He held a position at TUDelft in 2011-12, then he joined to the Control Systems (CS) Group at the Eindhoven University of Technology (TU/e). Currently, he is a Full Professor at the CS Group, TU/e and a Senior Researcher at the Systems and Control Laboratory, HUN-REN Institute for Computer Science and Control (SZTAKI) in Budapest, Hungary. He is Senior Editor of the IEEE Transactions on Control Systems Technology. His research interests are in identification and control of linear parameter-varying (LPV) and nonlinear systems, developing data-driven and machine learning methods with performance and stability guarantees for modeling and control, model predictive control and behavioral system theory. On the application side, his research focuses on advancing reliability and performance of precision mechatronics and autonomous robots/vehicles with nonlinear, LPV and learning-based motion control. He has received the TUDelft Young Researcher Fellowship Award in 2010, the VENI award of The Netherlands Organization for Scientific Research in 2011, the Starting Grant of the European Research Council in 2016 and the DCRG Fellowship of Mathworks in 2022.
\end{IEEEbiography}
\begin{IEEEbiography}[{\includegraphics[width=1in,height=1.25in,clip,keepaspectratio]{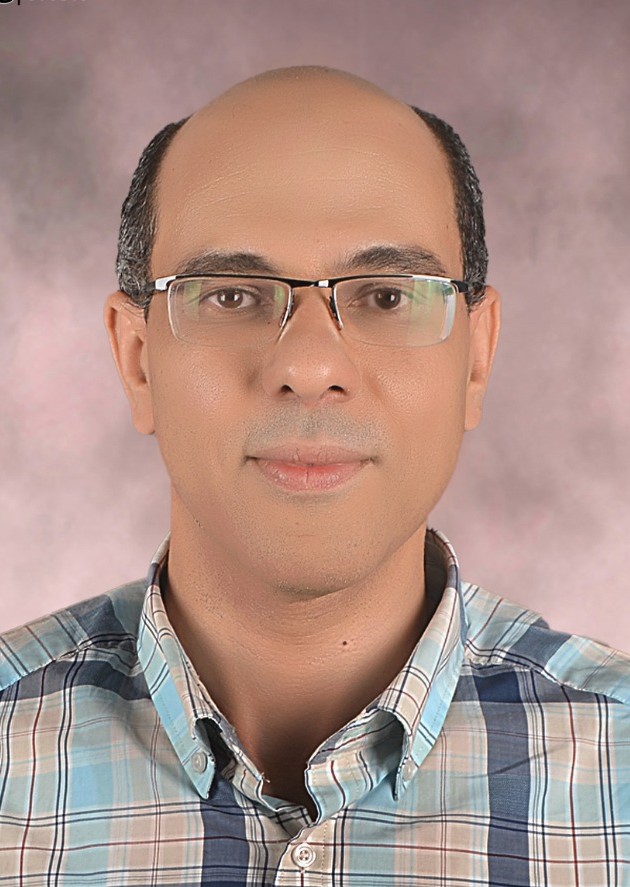}}]{Hossam S. Abbas} Hossam S. Abbas completed his B.Sc. and M.Sc. degrees in electrical engineering from Assiut University, Egypt, in 1997 and 2001, respectively, and his Ph.D. with focus on control systems from Hamburg University of Technology, Germany, in 2010. He was an Assistant Professor in the Electrical Engineering Department, Faculty of Engineering, Assiut University from 2010 to 2015, where he is an Associate Professor since 2015. 

He was a research fellow in Hamburg University of Technology, and Eindhoven University of Technology, the Netherlands, in 2011 and 2013, respectively. He was a senior researcher (Humboldt fellow) with the Institute for Electrical Engineering in Medicine, Universit\"at zu L\"ubeck, Germany, and with the Medical Laser Center in L\"ubeck, in 2017 and 2019, respectively. Currently, he is a senior scientist with the Institute for Electrical Engineering in Medicine.
\end{IEEEbiography}

\end{document}